\documentclass[aos]{imsart}
\RequirePackage{amsthm,amsmath,amsfonts,amssymb}
\RequirePackage[numbers]{natbib}
\RequirePackage[colorlinks,citecolor=blue,urlcolor=blue]{hyperref}
\RequirePackage{graphicx}

\usepackage[notextcomp]{kpfonts}
\usepackage{multirow}
\usepackage{enumitem} 

\startlocaldefs

\theoremstyle{plain}
\newtheorem{theorem}{Theorem}
\newtheorem{proposition}{Proposition}

\newtheorem{result}{Result}
\theoremstyle{remark}
\newtheorem{definition}{Definition}
\newtheorem{assumption}{Assumption}

\DeclareMathOperator*{\ran}{ran}
\DeclareMathOperator*{\spn}{span}
\DeclareMathOperator*{\argmin}{arg\,min}

\endlocaldefs

\begin{document}

\begin{frontmatter}
\title{Envelope-based partial least squares in functional regression}

\begin{aug}
\author[A]{\fnms{Minxuan}~\snm{Wu}}\ead[label=e1]{wuminxuan@ufl.edu},
\author[B]{\fnms{Joseph}~\snm{Antonelli}\ead[label=e2]{jantonelli@ufl.edu}}
\and
\author[C]{\fnms{Zhihua}~\snm{Su}\ead[label=e3]{z.sophia.su@gmail.com}}

\address[A]{wuminxuan@ufl.edu \printead[presep={,\ }]{e1}}

\address[B]{jantonelli@ufl.edu \printead[presep={,\ }]{e2}}

\address[C]{z.sophia.su@gmail.com\printead[presep={,\ }]{e3}}
\end{aug}

\begin{abstract}

In this article, we extend predictor envelope models to settings with multivariate outcomes and multiple, functional predictors. We propose a two-step estimation strategy, which first projects the function onto a finite-dimensional Euclidean space before fitting the model using existing approaches to envelope models. We first develop an estimator under a linear model with continuous outcomes and then extend this procedure to the more general class of generalized linear models, which allow for a variety of outcome types. We provide asymptotic theory for these estimators showing that they are root-$n$ consistent and asymptotically normal when the regression coefficient is finite-rank. Additionally we show that consistency can be obtained even when the regression coefficient has rank that grows with the sample size. Extensive simulation studies confirm our theoretical results and show strong prediction performance of the proposed estimators. Additionally, we provide multiple data analyses showing that the proposed approach performs well in real-world settings under a variety of outcome types compared with existing dimension reduction approaches. 

\end{abstract}


\end{frontmatter}


\newcommand{\suml}[1]{\sum\limits_{#1}}
\newcommand{\coorb}[1]{[#1]_{b}}
\newcommand{\coorc}[1]{[#1]_{c}}
\newcommand{\KL}{Karhunen-Lo\`eve}
\newcommand{\iX}{X_{i}}
\newcommand{\HX}{\mathcal{H}_{\bX}}
\newcommand{\Ho}{\mathcal{H}_{X_0}}

\newcommand{\inprod}[3][\Hc_{\bX}]{\langle#2,#3\rangle_{#1
	}}
\newcommand{\inprodi}[3][\mathcal{H}_{\iX}]{\langle#2,#3\rangle_{#1
	}}
\newcommand{\indep}{\perp\kern-6pt\perp}
\newcommand{\bX}{\mathbf{X}}
\newcommand{\union}[1]{\bigcup\limits_{#1}}
\newcommand{\HY}{\mathcal{H}_{Y}}
\newcommand{\dirctsuml}[1]{\bigoplus\limits_{#1}}
\newcommand{\Hp}{\mathcal{H}_{X^p}}
\newcommand{\Sc}{\mathcal{S}}
\newcommand{\U}{\mathcal{U}}
\newcommand{\V}{\mathcal{V}}

\newcommand{\Hi}[1]{\mathcal{H}_{X_{#1}}}
\newcommand{\bd}[1]{\boldsymbol{#1}}
\newcommand{\Rs}{\mathbb{R}}
\newcommand{\Ec}{\mathcal{E}}
\newcommand{\E}{\Ec}
\newcommand{\Hc}{\mathcal{H}}
\newcommand{\Bc}{\mathcal{B}}
\newcommand{\Ns}{\mathbb{N}}
\newcommand{\Xm}{X^{(m)}}
\newcommand{\Xcon}{X^{(c)}}
\newcommand{\HcXcon}{\Hc_{\Xcon}}
\newcommand{\Pcon}{P_{\HcXcon}}
\newcommand{\HXm}{\Hc_{\Xm}}
\newcommand{\SigmaX}{\Sigma_{\bX}}
\newcommand{\SigmaY}{\Sigma_{Y}}
\newcommand{\penvlp}{\Ec(B^*;\SigmaX)}
\newcommand{\Bcon}{B^{(c)}}
\newcommand{\Wc}{\mathcal{W}}
\newcommand{\cspan}{\overline{\text{span}}}
\newcommand{\TSigmaX}{\Tilde{\Sigma}_X}


\newcommand{\Cov}{\text{Cov}}

\newcommand{\Uc}{\mathcal{U}}
\newcommand{\bY}{\bd{Y}}

\newcommand{\HXmx}{\HX^{(m_x)}}
\newcommand{\HYmy}{\HY^{(m_y)}}
\newcommand{\Xmx}{\bX^{(m_x)}}
\newcommand{\Ymy}{Y^{(m_y)}}
\newcommand{\epsilonmy}{\epsilon^{(m_y)}}
\newcommand{\Bmymx}{B_{m_y,m_x}}
\newcommand{\emy}{\epsilon^{(m_y)}}

\newcommand{\evmy}{\Tilde{\epsilon}^{(m_y)}}
\newcommand{\Yvmy}{\Tilde{Y}^{(m_y)}}
\newcommand{\Xvmx}{\Tilde{\bX}^{(m_x)}}
\newcommand{\Bmmymx}{\Tilde{B}_{m_y,m_x}}
\newcommand{\Bhatmymx}{\hat{B}_{m_y,m_x}}
\newcommand{\Bhatmmymx}{\hat{\bd{\beta}}_{m_y,m_x}}

\newcommand{\Bhatmmx}{\hat{\bd{\beta}}_{m_x}}
\newcommand{\Bmmx}{\Tilde{B}_{m_x}}
\newcommand{\Bhatmx}{\hat{B}_{m_x}}

\newcommand{\emymx}{\epsilon^{(m_y,m_x)}}
\newcommand{\evmymx}{\Tilde{\epsilon}^{(m_y,m_x)}}

\newcommand{\Var}{\text{Var}}



\section{Introduction}

Functional data has increased in popularity with advancements in modern technology, spanning numerous fields such as image or signal processing, climate science, machine learning, speech recognition, and more (\cite{ramsay_silverman_2010}, \cite{wang_review_FDA_2016}, \cite{wang2024review}).  Functional data includes curves and images, which are commonly viewed as infinite-dimensional random vectors in a functional space (\cite{hsing2015theoretical},\cite{wang_review_FDA_2016}). Similar to more traditional settings, the basic goals of functional data analysis (FDA) still include regression, classification, and descriptive analyses, which are the same as those of multivariate data. For this reason many corresponding methods have been extended to function-valued data, however, functional data are intrinsically infinite-dimensional, which poses additional challenges including infinite/high-dimensionality of observations and parameters as well as misalignment (\cite{gertheiss2024functional}).


In this paper, we extend the dominant method in the area of chemometrics (\cite{mehmood2016diversity}, \cite{wold2001pls}), partial least squares (PLS), from multivariate regression to functional regression. PLS was first introduced in \cite{wold1966estimation} and can be applied to address collinearity in high-dimensional regression problems, which is broadly used in other applied science such as bioinformatics (\cite{boulesteix2007partial}, \cite{nguyen2002tumor}) and social science (\cite{hulland1999use}, \cite{mateos2011partial}). When there is a high degree of collinearity in the data, PLS frequently outperforms ordinary least squares and principle component regression. There exists some work tailoring PLS to functional data analysis in different contexts, (\cite{AGUILERA201680}, \cite{beyaztas2020function}, \cite{Delaigle2012FPLS}, \cite{escabias_aguilera_valderrama_2007}, \cite{preda2007pls}, \cite{Reiss01092007}), however, due to iterative nature and inexplicit form of the algorithm for PLS and its functional extension, it is very challenging to derive the theoretical properties for them (\cite{Delaigle2012FPLS}). To the best of our knowledge, there is limited understanding of the theoretical properties of PLS and functional PLS within the framework of a well-defined population model because of its iterative nature. Some theoretical results, such as those in \cite{preda2007pls}, which build on \cite{höskuldsson_1988}, include convergence of the PLS approximation but without proof or convergence rates. While more explicit consistency results are developed in \cite{Delaigle2012FPLS} using an alternative formulation, a comprehensive theoretical understanding remains limited. Therefore, it is of particular interest not only to propose a functional PLS but also to present theoretical results, such as consistency, convergence rates, and asymptotical normality. This not only provides theoretical justification for using these models, but also opens the door for inference on regression coefficients and prediction intervals. 

The envelope model is a recently developed dimension reduction technique using the reducing subspaces of the covariance matrix, which was first introduced by \cite{cook_li_chiaromonte_2010}. Later, this concept was developed to more general contexts, such as partial least squares \cite{cook_helland_su_2013}, generalized linear models (\cite{cook_2015_foundations}, sparse partial least squares (\cite{zhu_2020_envelopebased}), partial partial least squares (\cite{park_su_chung_2022}), scalar-on-function regression \citep{zhang_wang_wu_2018} and function-on-function regression (\cite{Su2022}), among many others. Of particular relevance for this manuscript is that the predictor envelope was shown to be the fundamental goal of PLS in \cite{cook_helland_su_2013}. This connection between envelope models and PLS makes it possible to derive tractable theoretical properties for PLS within a traditional Fisherian framework of well-defined population parameters; see \cite{cook2024partial} for an extensive discussion. Our manuscript develops results for a functional extension of PLS, which exploits this connection and is based on a functional version of the predictor envelope. 

There is a gap in the literature on how to perform functional PLS with provable guarantees, which we aim to fill in this manuscript. We provide a framework for functional predictor envelope models and propose two novel functional versions of PLS: (1) functional envelope-based
PLS (FEPLS), which lies in the scope of functional linear regression, and (2) generalized functional envelope-based PLS (GFEPLS), which adapts to more flexible response distributions. We derive the theoretical properties of the proposed estimators by establishing that they are root-$n$ consistent when the regression coefficient is finite-rank, and consistent even when an infinite-dimensional functional envelope occurs. Moreover, we also construct pointwise confidence and prediction intervals based on an asymptotic normal approximation approximation. Through both simulation studies and multiple data analyses, we show strong performance in terms of prediction error and dimension reduction.

\section{Review of multivariate PLS and envelopes}
In this section, we provide a brief overview of the predictor envelope in the multivariate setting and its connection to partial least squares. Let $\bY\in\Rs^{d_y}$ be the response vector and let $\bX\in\Rs^{d_x}$ be the predictor vector. Consider the multivariate linear model
\begin{equation*}
\bY=\bd{\alpha}+\bd{\beta}\bX+\epsilon,
\end{equation*}
where $\bd{\alpha}\in\Rs^{d_y}$ is the intercept and $\bd{\beta}\in\Rs^{d_y\times d_x}$ is the regression coefficient matrix. The error term $\epsilon\in\Rs^{d_y}$ has mean 0 and covariance matrix $\Sigma_\epsilon$,  the random vector $\bX\in\Rs^{d_x}$ is independent of $\epsilon$ and has mean $0$ and covariance matrix $\SigmaX$.

\subsection{PLS} 
Partial least squares is a statistical method that reduces the number of predictors to a fixed number of linear combinations of predictors, $\bd{W}^T\bX$, where $\bd{W}\in\Rs^{d_x\times u}$ and $u\leq d_x$. There are two common iterative algorithms for finding $\boldsymbol{W}$: NIPALS (\cite{wold_1975_NIPALS}) and SIMPLS (\cite{dejong_1993}). Both algorithms sequentially produce the $k$th column vector, $\bd{w}_k$, of $\bd{W}$, for $k<u$. The two algorithms have the same objective function but they differ in their constraints, however, for simplicity we focus attention here only on the SIMPLS algorithm. Letting $\bd{W}_k=(\bd{w}_1,\dots,\bd{w}_k)$, the $(k+1)$ column vector $\bd{w}_{k+1}$ can be obtained as:
\begin{equation*}
    \bd{w}_{k+1}=\arg\max_{\bd{w}}\bd{w}^T\Sigma_{\bd{XY}}\Sigma_{\bd{XY}}^T\bd{w},\ \text{subject to } \bd{w}^T\SigmaX \bd{W}_k=0\text{ and }\bd{w}^T\bd{w}=1,
\end{equation*}
where $\Sigma_{\bd{XY}}$ is the cross-covariance matrix between $\bX$ and $\bY$. This process is iterated until the desired number of components is found, where the number of the components is typically selected by cross-validation. The PLS estimator of $\bd{\beta}$ is $\hat{\bd{\beta}}_{PLS}=\bd{W}(\bd{W}^T\bd{S}_X\bd{W})^{-1}\bd{W}^T\bd{S}_{XY}$, where $\bd{S}_X\in\Rs^{d_x\times d_x}$ is the sample covariance matrix of $\bd{X}$, and $\bd{S}_{XY}\in\Rs^{d_x\times d_y}$ is the sample covariance matrix of $\bd{X}$ and $\bd{Y}.$

\subsection{Envelope-based PLS}

\subsubsection{Multivariate predictor envelope linear model}
\label{subsec: MPELM}

Consider $\Sc\subseteq\Rs^{d_x}$ and let $\Sc^{\perp}$ denote its orthogonal complement. Let $P_\Sc$ denote the projection matrix onto $\Sc$ and $Q_\Sc=I-P_\Sc$, be the projection matrix onto $\Sc^{\perp}$. Assume that $\Sc$ satisfies the following conditions: (i) $\Cov(\bd{Y},Q_{\Sc}X|P_{\Sc}X)=0$, and (ii) $\Cov(P_{\Sc}\bX,Q_{\Sc}\bX)=0$. These conditions ensure that there is no direct linear relationship between $\bY$ and $Q_\Sc\bX$ when $P_\Sc\bX$ is known, and that there is no marginal linear relationship between $P_\Sc\bX$  and $Q_\Sc\bX$. Thus, $P_\Sc\bX$ and $Q_\Sc\bX$ are referred to as the (linearly) material part and immaterial part of $\bX$, respectively. It can be shown that these conditions are equivalent to stating that $\Sc$ is a reducing subspace of $\SigmaX$ that contains $\Uc$, where $\Uc=\spn(\bd{\beta}^T)$ (\cite{cook_helland_su_2013}). The reducing subspace of $\SigmaX$ is an invariant subspace of $\SigmaX$ whose orthogonal complement is also an invariant subspace of $\SigmaX$. The predictor envelope, denoted by $\Ec(\bd{\beta}^T;\SigmaX)$, is defined as the smallest reducing subspace of $\SigmaX$ that contains $\Uc$. Let $u$ denote the dimension of $\Ec(\bd{\beta}^T;\SigmaX)$.

Let $\Phi\in \Rs^{d_x\times u}$ be an orthonormal basis of $\Ec(\bd{\beta}^T;\SigmaX)$, and let $\Phi_0\in\Rs^{d_x\times (d_x-u)}$ be an orthonormal completion of $\Phi$, where a matrix $A$ is called a basis matrix of a subspace $\mathcal{S}$ if the columns of $A$ form a basis of $\mathcal{S}$. Since $\mathcal{U}$ is a subspace of $\Ec(\bd{\beta}^T;\SigmaX)$, we can write $\bd{\beta}^T=\Phi\bd{\eta}$. The coordinate form of the multivariate predictor envelope linear model (MPELM) is:
\begin{equation*}
    Y=\alpha+\eta^T\Phi^T(X-\mu_X)+\epsilon,\qquad \SigmaX={\Phi}\Delta{\Phi}^T+{\Phi}_0\Delta_0{\Phi}_0^T,
\end{equation*}
where $\Delta=\Phi^T\SigmaX\Phi$ and $\Delta_0=\Phi_0^T\SigmaX\Phi_0$. Let $\bd{S_Y}$ denote the sample covariance matrix of $\bd{Y}$ and $\bd{S_{Y|X}}=\bd{S_X}-\bd{S_{XY}}\bd{S}^{-1}_{\bd{Y}}\bd{S}_{\bd{XY}}^{T}$ be the sample covariance matrix of $\bX$ given $\bY$. The maximum likelihood estimator of $\Phi$ can be obtained by
\begin{equation}
    \widehat{\Phi}=\argmin_{\bd{G}}( \log|\bd{G}^T\bd{S}_{X|Y}\bd{G}|+\log|\bd{G}^T\bd{S}_{X}^{-1}\bd{G}|),
    \label{eq: obj func of MPELM}
\end{equation}
where the minimization is over all semiorthogonal matrices $\bd{G}\in\Rs^{d_x\times u}$. The objective function in (\ref{eq: obj func of MPELM}) is invariant under right orthogonal transformation, so the minimization is over the Grassmann manifold and the solution is not unique. In \cite{cook_helland_su_2013}, it was shown that $\spn(\bd{W}_u)=\Ec(\bd{\beta}^T;\SigmaX)$ and the SIMPLS algorithm produces a root-$n$-consistent estimator of the projection onto the predictor envelope $\Ec(\bd{\beta}^T;\SigmaX)$. The SIMPLS algorithm is therefore an approach to estimate the predictor envelope $\Ec(\bd{\beta}^T;\SigmaX)$. Furthermore, it also showed that the likelihood-based envelope estimator generally achieves greater efficiency compared to other PLS methods.

\subsubsection{Generalized envelope linear model} 
\label{subsec: GMELM}
The core idea of the predictor envelope was further developed in \cite{cook_2015_foundations} for a more general regression setting. Specifically, the envelope estimator for generalized linear models (GLM) with a canonical link function was discussed, which can be adapted to a general link function (see \cite{zhu_2020_envelopebased}). Considering that the distribution of $Y$ belongs to a natural exponential family, the coordinate form of generalized envelope linear model (GMELM) is 
\begin{equation*}
    \begin{split}
        &log(f(Y|\theta))=Y\theta-b(\theta)+c(y),\qquad \theta(\mu)=(b')^{-1}(g^{-1}(\mu))\\
        &\mu=\alpha+\bd{\eta}^T{\Phi}^T\bX,\qquad\SigmaX={\Phi}\Delta{\Phi}^T+{\Phi}_0\Delta_0{\Phi}_0^T,
    \end{split}
\end{equation*}
where $\Delta=\Phi^T\SigmaX\Phi$ and $\Delta_0=\Phi_0^T\SigmaX\Phi_0$. Also, $\theta$ is the natural parameter ($\theta(\cdot)$ denotes $(b')^{-1}(g^{-1}(\cdot))$), $b(\cdot)$ is the cumulant function, $g(\cdot)$ is a monotonic link function, and $c(\cdot)$ is some specific function.  Moreover, let $\Phi$ be an orthonormal basis of $\Ec(\bd{\beta}^T;\SigmaX)$ and $\Phi_0$ be a completion of $\Phi$. The estimators of ($\alpha,\bd{\eta},\Phi$) are obtained by minimizing 
\begin{equation}
        L_n(\alpha,\bd{\eta},{\Phi})=-\frac{2}{n} \suml{i=1}^n D(\alpha+\bd{\eta}^T{\Phi}^T{\bX_i})+\log|{\Phi}^T{S_X}{\Phi}|+\log|{\Phi}^T{S_X}^{-1}{\Phi}|,
    \label{eq: obj func of GMELM}
\end{equation}
where $D(\cdot)=\mathcal{C}(\theta(\cdot))$ and $\mathcal{C}(\nu)=Y\nu-b(\nu)$.

This objective function (\ref{eq: obj func of GMELM}) can be optimized through a two-step alternating optimization process. First, given ${\Phi}$,  (\ref{eq: obj func of GMELM}) becomes a function of ($\alpha,{\eta}$), which can be optimized using the Fisher scoring method or other methods typically employed for fitting GLMs. Second, given $\alpha$ and ${\eta}$, one can minimize (\ref{eq: obj func of GMELM}) with respect to $\Phi$, which involves optimization over the Grassmann manifold. A fast algorithm for Grassmann manifold optimization was proposed in \cite{cook_forzani_su_2016}, which employs a novel reparameterization technique to transform the original problem into an unconstrained non-manifold optimization problem. The unconstrained non-manifold optimization problem can then be addressed through an iterative descent method applied to each row of $\Phi$. 
The aforementioned two steps are recursively implemented until a convergence criterion is met. A comprehensive algorithm for the predictor envelope in the logistic regression setting is provided in the Appendix.

\section{Notations and definitions}
This section introduces the notations and definitions used throughout the rest of paper. Let $L^2[a,b]$ denote all $L^2$ integrable functions on $[a,b]$. With an inner product $\inprod[L^2]{x_1}{x_2}:\int_a^bx_1(t)x_2(t)dt$, $L_2[a,b]$ forms a separable Hilbert space. More formal definitions and results regarding random functions can be found in \cite{kokoszka_reimherr_2017}. We use $\Hc$ as a general notation for the Hilbert space,  $\inprod[\Hc]{\cdot}{\cdot}$ to denote the inner product on $\Hc$ and $||\cdot||_{\Hc}$ to denote the norm induced by the inner product. 

\subsection{Functional envelope}
We will formally state the novel definition, introduced by \cite{Su2022}, for expectation and covariance of a random element in a Hilbert space. Based on them, we can then define the envelope on a general Hilbert space, leading to the functional envelope on $L^2[a,b]$.

\begin{definition}
    For a random element $X\in\Hc$, the \textbf{expectation} of $X$, denoted by $EX$, is defined as a linear functional $l_X:\Hc\to\Rs$ such that $\forall f\in\Hc$, 
    \begin{equation*}
        E(\inprod[\Hc]{X}{f})=\inprod[\Hc]{l_X}{f}.
    \end{equation*}
    \label{def_expectation}
\end{definition}

The tensor product of $f\in\Hc_1$ and $g\in\Hc_2$, denoted by $f\otimes g$, is defined as the linear operator $L_{f\otimes g}: \Hc_2\to\Hc_1$ such that, for any $c\in\Hc_2$, $L_{f\otimes g}(c)=f\inprod[\Hc]{g}{c}$. Let $\Bc(\Hc_1,\Hc_2)$ denote the set of all bounded linear operators from $\Hc_1$ to $\Hc_2$, and let $\Bc(\Hc)$ denote $\Bc(\Hc,\Hc)$. For any $B\in\Bc(\Hc_1,\Hc_2)$, let $B^*\in\Bc(\Hc_2,\Hc_1)$ denote the adjoint operator of $B$.

\begin{definition}
     For a random element $X\in\Hc$, the \textbf{second moment} of $X$, denoted by $E(X\otimes X) $, is defined as the expectation of $X\otimes X$, which is a bounded linear operator $B\in\Bc(\Hc)$ such that for any $f,g\in\Hc$, 
    \begin{equation*}
        E(\inprod[\Hc]{f}{(X\otimes X)g})=\inprod[\Hc]{f}{B g}.
    \end{equation*}
\end{definition}
Following the previous definitions, one can define the variance operator of $X$ as $E((X-EX)\otimes (X-EX))$, and denote it by $\Var(X)$. The existence of expectation and second moments has been discussed in \cite{Su2022} and they are well-defined. Next, we will formally state the definition of the $A-$envelope on a Hilbert space for a self-adjoint operator $A$.

\begin{definition}
    Let $\mathcal{U}$ be a subspace of $\mathcal{H}$ and $A:\ \mathcal{H}\to\mathcal{H}$ be a self-adjoint operator. The \textbf{$A$-envelope} of $\mathcal{U}$, denoted by $\Ec(\mathcal{U};A)$, is defined as the intersection of all reducing subspaces of $A$ that contain $\mathcal{U}$. Mathematically,
    \begin{equation*}
        \Ec(\mathcal{U};A)=\cap\{\mathcal{S}:\mathcal{S}\in\text{Lat}_{\mathcal{U}}(A)\} ,
    \end{equation*}
    where the lattice $\text{Lat}_{\mathcal{U}}(A)$ includes all reducing subspaces of $A$ that contain $\mathcal{U}$. 
    \label{def: generalized envelope}
\end{definition}

For any $B\in\Bc(\Hc)$, let ran($B$) denote the range of $B$. We abbreviate $\Ec(\overline{\text{ran}}(B);A)$ as $\Ec(B;A)$, where $\overline{\text{ran}}(B)$ denotes the closure of $\text{ran}(B)$. Since $L^2[a,b]$ is a Hilbert space, Definition \ref{def: generalized envelope} is applicable for a self-adjoint operator $A$ on $L^2[a,b]$, also referred to as the functional envelope. Compared to the envelope defined in a Euclidean space, $\Ec(\mathcal{U};A)$ can potentially be infinite-dimensional, even when $\mathcal{U}$ is finite-dimensional. This makes functional envelope-based estimation significantly more challenging.

\section{Functional predictor envelope linear model}
\subsection{Functional predictor envelope linear model}
Consider $\Hc_{X_i}=L^2[a_i,b_i]$ for $i=1,2,\dots, p$ and $\HY$ is a separable Hilbert space. The general framework for the functional linear model with $p$ predictors is given by:
 \begin{equation*}
     Y=\alpha+\suml{i=1}^pB_iX_i(\cdot)+\epsilon,
 \end{equation*}
where $X_i(\cdot)\in\Hc_{X_i}$ for $i=1,2,\dots, p$, and $Y\in\HY$. The error term $\epsilon\in\HY$ is a Gaussian process with mean 0 and variance operator $\Sigma_{\epsilon}$ and is independent of $(X_1(\cdot),...,X_p(\cdot))$, and $B_i\in\Bc(\Hc_{X_i},\HY)$ is compact. The functions $X_i(\cdot)\in\Hc_{X_i}$ are Gaussian random functions defined over intervals, with mean 0 and compact covariance operator $\Sigma_{X_i}$. For simplicity, we use $X_i$ as an abbreviation for $X_i(\cdot)$. The domains of the predictors may differ, but by a change of variables, we change their domains all to $[0,1]$. We further assume $\Hc_{X_i}$ is a closed subspace of $L^2[0,1]$, making it a separable Hilbert space. Without loss of generality, we always assume $E(Y)=0$ and $EX_i=0$, and therefore $\alpha=0$.

Let $\bX(\cdot)=(X_1(\cdot),X_2(\cdot),...,X_p(\cdot))^T$ and $\Hc_{\bX}=\varprod\limits_{i=1}^p\mathcal{H}_{X_i}$, where $\varprod$ represents the Cartesian product. Then, $\bX(\cdot)\in\Hc_{\bX}$ is a real, vector-valued random function on $[0,1]$, and  $\Hc_{\bX}$ remains a separable Hilbert space. Let $B: \Hc_{\bX}\to \HY,\ f\to \sum_{i=1}^{p}B_{i}f_{i}$, which is still a compact bounded linear operator. Thus, for multiple functional predictors, one can still write the functional linear Model as 
\begin{equation}
    Y=B\bX(\cdot)+\epsilon,
    \label{eq: FLM}
\end{equation}
where $\bX(\cdot)$ is a vector-valued Gaussian random function with mean 0 and compact covariance operator $\Sigma_{X}$, and $Y$ is a Gaussian random element in $\HY$ with mean 0. We refer to (\ref{eq: FLM}) as the functional linear model (FLM).

Let $\penvlp$ denote the predictor envelope of (\ref{eq: FLM}) and $P_{\Ec}$ denote the projection onto $\penvlp$. We denote $Q_{\Ec}=I-P_{\Ec}$, where $I$ is the identity operator. The following proposition defines key properties of the functional predictor envelope. 
\begin{proposition}[Properties of the functional predictor envelope]
    The following holds:
    \begin{longlist}
        \item $\penvlp$ is a reducing subspace of $\SigmaX$.
        \item $P_{\Ec}\bX$ is independent of $Q_{\Ec}\bX$.
        \item $Y$ is independent of $Q_{\Ec}\bX$.
        \item $\SigmaX=P_{\Ec}\SigmaX P^*_{\Ec}+Q_{\Ec}\SigmaX Q^*_{\Ec}$.
        \item \label{prop item: basis} There exists an orthonormal basis $E_x=\{e_{xj}:j\in \Ns\}$ of $\HX$ and an index set $j$ such that $e_{xj}$ is the eigenfunction of $\SigmaX$, ($j\in\Ns$), and $\{e_{xj}:j\in J\}$ is an orthonormal basis of $\penvlp$.
    \end{longlist}
    \label{prop: properties of FPE}
\end{proposition}

Then, the functional predictor envelope linear model (FPELM) with a functional predictor envelope $\penvlp$ is
\begin{equation}
    Y=\eta^* P_{\Ec} \bX + \epsilon,\qquad \SigmaX=P_{\Ec}\SigmaX P^*_{\Ec}+Q_{\Ec}\SigmaX Q^*_{\Ec},
    \label{eq: FPELM}
\end{equation}
where $\eta^*\in\Bc(\penvlp,\HY)$ satisfies that $B=\eta^* P_{\Ec}$. Since $Y$ is independent of $Q_{\Ec}\bX$, we also refer $P_{\Ec} \bX$ and $Q_{\Ec} \bX$ to as the material part and immaterial part, respectively. In order to characterize the functional predictor envelope we need the following assumption.

\begin{assumption}[Finite-rank]
    Assume $B$ is finite-rank.
    \label{assump: finite-rank operator}
\end{assumption}

Assumption \ref{assump: finite-rank operator} imposes a sparse structure on the regression operator, making sufficient dimension reduction possible with a finite number of components. While this assumption may not always hold, it is reasonable to assume that a finite number of components capture the majority of the information in $\bX$, which still leads to sound performance. Note that although $B$ exhibits sparsity, $\penvlp$ and $\HY$ can still be infinite-dimensional. The finite-dimensional predictor envelope, along with the finite-dimensional response envelope, was discussed in \cite{Su2022} and is referred to as the eigen-sparse envelope. The FPELM with an eigen-sparse predictor envelope guarantees that the finite-rank assumption holds, while the finite-rank assumption does not necessarily an eigen-sparse predictor envelope. For instance, $\ran(B^*) = \spn(\{t\})$ and the eigenfunctions of $\SigmaX$ are the Fourier basis with different eigenvalues. In this case, $B$ is finite-rank, but $\penvlp$ is infinite-dimensional.

Next, we characterize the functional predictor envelope under a special orthonormal basis of $\Hc_{\bX}$. By the spectral decomposition of a compact self adjoint operator $\Sigma_{X}$, the eigenfunctions of $\SigmaX$ can form an orthonormal basis of $\HX$, and we denote this orthonormal basis by $E_x=\{e_{xj}:j\in\Ns\}$. Let $E_y=\{e_{yi}:i\in\Ns\}$ be an arbitrary orthonormal basis of $\HY$ and $y_i=\inprod[\HY]{e_{yi}}{Y}$.

\begin{proposition}
            Suppose Assumption \ref{assump: finite-rank operator} holds.  Then, (\ref{eq: FPELM}) can also be written as:
    \begin{equation*}
        \suml{i\in\Ns}y_ie_{y_i} = \suml{i\in \Ns}\left(\suml{j\in J}\beta_{ij}x_{j} \right)e_{yi}+\epsilon,\qquad \SigmaX = \suml{j\in J} \sigma_{xj}^2e_{xj}\otimes e_{xj} + \suml{j\notin J} \sigma_{xj}^2e_{xj}\otimes e_{xj},
    \end{equation*}
    where $J\in\Ns$ is an index set and
    $\spn\{e_{xj}:j\in J\}=\penvlp$.

    \label{prop: FPELM basis expansion}
\end{proposition}

With $E_x$ and $E_y$, Proposition \ref{prop: FPELM basis expansion} expresses the FPELM in terms of infinite-dimensional vectors $\{y_i:i\in\Ns\}$ and $\{x_j:j\in\Ns\}$, as well as infinite-dimensional matrices $\{\beta_{ij}:i\in\Ns,\ j\in J \}$ and $\{\sigma_{xj}:j\in \Ns\}$. When considering a vector or scalar response, Assumption \ref{assump: finite-rank operator} holds. The first part of Proposition \ref{prop: FPELM basis expansion} can be expressed in terms of finite-dimensional vectors and the regression coefficient matrix, but the infinite-dimensional functional covariance operator structure remains. 

Since $\ran(B^*)$ is finite-dimensional under Assumption \ref{assump: finite-rank operator}, then there exists a finite-dimensional $\HcXcon\subset\HX$ such that $\ran(B^*)\subseteq\HcXcon$. Let $\Pcon$ denote the projection of $\Hc_{\bX}$ onto $\HcXcon$, and $\Xcon=P_{\HcXcon}X$ denote the projected $X$ on $\HcXcon$. Since $\Xcon$ includes all the material part of $X$, then ($Y,\Xcon$) still follows the FELM: 
\begin{equation}
    Y=\Bcon\Xcon + \epsilon,\ \Sigma_{\Xcon}=\suml{j\in J'}\sigma^2_{xj}{e}'_{xj}\otimes {e}'_{xj} + \suml{k\in K}\sigma^2_{xk} {e}'_{xk}\otimes {e}'_{xk},
    \label{eq: projected FPELM}
\end{equation}
where $\Bcon:\ \Hc_{\Xcon}\to\HY$ satisfies that $f\in\Hc_{\Xcon}$, $\Bcon f= Bf$, and $\Sigma_{\Xcon}$ is the covariance operator of $\Xcon$. Moreover, $\{e'_{xj}:j\in J'\}\cup \{e'_{xk}:k\in K\}$ are eigenfunctions of $\Sigma_{\Xcon}$ and $J'$ and $K$ are index sets. However, the predictor envelope of (\ref{eq: projected FPELM}), $\Ec(({\Bcon})^*;\Sigma_{\Xcon})$, is different from (\ref{eq: FPELM}). One significant difference is that $\Ec(({\Bcon})^*;\Sigma_{\Xcon})$ is always finite-dimensional while $\penvlp$ may not be.

Consider that $\bd{b}_{m_x}=(b_{m_x,1},...,b_{m_x,m_x})$ are some orthonormal basis functions of $\HX$, and  $\bd{c}_{m_y}=(c_{m_y,1},,...,c_{m_y,m_y})$ are orthonormal basis functions of $\HY$, or an orthonormal basis of $\HY$ if $\HY$ is finite-dimensional. Let $\HXmx$ denote the $\text{span}(\bd{b}_{m_x})$ and $\HYmy$ denote the $\text{span}(\bd{c}_{m_y})$. We use $P_{\Hc}$ to denote the projection onto $\Hc$ for any $\Hc$. Let $\Xmx=P_{\HXmx}\bX$ and $\Ymy=P_{\HYmy}Y$. If we assume there exists $m_x,m_y\in\Ns$ such that $\ran(B^*)\subseteq\HXmx$ and $\ran(B)\subseteq\HYmy$ (since $\ran(B)$ is finite-rank under Assumption \ref{assump: finite-rank operator}), it follows that we can set $\HcXcon=\HXmx$, thereby inheriting the notations from (\ref{eq: projected FPELM}). Furthermore, the following proposition holds.

\newcommand{\evxj}{\Tilde{e}_{xj}}


\begin{proposition}
    Suppose that $\ran(B^*)\subseteq\HXmx$. Then, for any $\bd{c}_{m_y}$, $(\Xmx,\Ymy)$ still follows the FPELM:
    \begin{equation}
        \begin{split}
            &\Ymy=\Bmymx\Xmx+\emy,\\
            &\Sigma_{\Xmx}=\suml{j\in J}\sigma_{xj}^2 P_{\Xmx}e_{xj}\otimes (P_{\Xmx}e_{xj}) + \suml{j\notin J}\sigma_{xj}^2P_{\Xmx}e_{xj}\otimes (P_{\Xmx}e_{xj}),
        \end{split}
        \label{eq: projected FPELM mx my}
    \end{equation}
    where $\Bmymx:\HXmx\to\HYmy $ satisfies that for any $f\in\HXmx$, $\Bmymx f=Bf$ and $\epsilonmy=P_{\HYmy}\epsilon\in\HYmy$. With $\bd{b}_{m_x}$ and $\bd{c}_{m_y}$, (\ref{eq: projected FPELM mx my}) can be written as a MPELM:
    \begin{equation}
        \Yvmy=\Bmmymx\Xvmx+\evmy,\qquad \Tilde{\Sigma}_{\Xvmx}=\suml{j\in\Ns}\sigma_{xj}^2 \evxj \evxj^T,
        \label{eq: projected MPELM}
    \end{equation}
    where $\Yvmy\in\Rs^{m_y}$ and $\evmy\in\Rs^{m_y}$ are the coordinates of $\Ymy$ and $\emy$ with respect to $\bd{c}_{m_y}$ respectively, and 
 $\Tilde{e}_{xj}^{(m_x)}\in\Rs^{m_x}$ and $\Xvmx\in\Rs^{m_x}$ are the coordinates of $P_{\HXmx}e_{xj}$ and $\Xmx$ with respect to $\bd{b}_{m_x}$ respectively.
    \label{prop: FPELM and MPELM}
\end{proposition}


Proposition \ref{prop: FPELM and MPELM} not only shows that a sufficient dimension reduction is possible, but also shows a one-to-one relation between the projected FPELM and MPELM. It is worth noting that (\ref{eq: FPELM}) and (\ref{eq: projected FPELM mx my}) have different predictor envelope structures, but (\ref{eq: projected MPELM}) and (\ref{eq: projected FPELM mx my}) have the same envelope structure under a set of orthonormal basis functions of $\HXmx$. The difference between (\ref{eq: FPELM}) and (\ref{eq: projected FPELM mx my}) could be substantial, as (\ref{eq: FPELM}) could have an infinite-dimensional functional envelope, but (\ref{eq: projected FPELM mx my}) cannot and there is an isomorphism between it and a MPELM.

\subsection{Estimation\label{subsec: estimation in FPELM}} 

The functional PLS can be constructed based on the functional predictor envelope. The infinite-dimensional nature of the functional predictor envelope obstructs a direct extension of classical estimation of envelope-based PLS. Proposition \ref{prop: FPELM and MPELM}, however, inspires a two-step procedure to construct the functional envelope-based PLS (FEPLS) estimator. 

The first step is the coordinate estimation of the projection, which requires choosing suitable orthonormal basis functions, $\bd{b}_{m_x}$ and $\bd{c}_{m_y}$. The common choices of basis functions include natural spline functions, spline bases, radial basis functions, and many others. The orthonormal basis functions can then be obtained by orthogonalizing these bases; for instance, \cite{Su2022} discusses methods for orthogonalization. The coordination estimation of $\Xvmx$ and $\Yvmy$ can be obtained via ordinary least squares (OLS), though they can also be obtained by penalized methods such as the least absolute shrinkage and selection operator (LASSO, \cite{tibshirani1996regression}) or ridge regression. In practice, to keep the material part, it might be preferable to choose a large set of basis functions.

The second step is to fit a MPELM for $(\Xvmx,\Yvmy)$, where the dimension for the envelope is chosen by the Bayesian Information Criterion (BIC). The estimation of MPELM involves optimization over the Grassmann manifold, and there is an existing R package \textit{Renvlp} in CRAN (see \cite{Renvlp}) that provides estimation for the MPELM. The prediction procedures are first obtaining the predicted $\Yvmy$ and then transforming it into a function. More details regarding estimation and prediction are included in Appendix B.

\subsection{Theoretical properties of FEPLS estimators} 
\label{subsec: theoretical results of FEPLS}  

Let $\Bhatmmymx\in\Rs^{m_y\times m_x}$ denote the MPELM estimator of $\Bmmymx$, and $\Bhatmymx=\suml{i=1}^{m_y} \suml{j=1}^{m_x}(\Bhatmmymx)_{i j}c_{m_y,i}\otimes b_{m_x,j}$ denote FEPLS estimator corresponding to $\Bhatmmymx$. First, we will show that under Assumption \ref{assump: finite-rank operator}, $\Bhatmymx$ is root-$n$-consistent for $B$ in the sense that its asymptotic variance is finite. Second, in a more general setting without Assumption \ref{assump: finite-rank operator}, where $\ran(B^*)$ can be infinite-dimensional, we demonstrate how a consistent sequence of FEPLS estimators for $B$ can be obtained using the two-step procedure described earlier. Let $\text{vec}(\cdot)$ denote the operator that stacks the columns of a matrix into a column vector. For the MPELM in (\ref{eq: projected MPELM}), the following proposition shows root-$n$-consistency of $\Bhatmmymx$.

\begin{proposition}
    Suppose Assumption \ref{assump: finite-rank operator} holds, $\ran(B^*)\subseteq\HXmx$, and $\ran(B)\subseteq \HYmy$. Then we have 
    \begin{equation*}
        \sqrt{n}(\text{vec}(\Bhatmmymx)-\text{vec}(\Bmmymx))\overset{d.}{\to}N(0,V_{mpelm}),
    \end{equation*}
    where $V_{mpelm}$ is the asymptotic variance of $\text{vec}(\Bhatmmymx)$.
    \label{prop: asymp normality of projected MPELM}
\end{proposition}

The expression of $V_{mpelm}$ requires additional notations and is included in the Appendix A as well as the proof. Notably, $\text{vec}(\Bhatmmymx)$ is asymptotically more efficient than or as efficient as  the maximum likelihood estimator of $\Bmmymx$.

\begin{theorem}
    Suppose Assumption \ref{assump: finite-rank operator} holds, $\ran(B^*)\subseteq\HXmx$, and $\ran(B)\subseteq \HYmy$. Then, we have
    \begin{equation*}
        nE||\Bhatmymx-B||^2_{op}<\infty, \text{as } n\to\infty,
    \end{equation*}
    where $||\cdot||_{op}$ denote the operator norm.
    \label{thm: root n consistency for FEPLS}
\end{theorem}

From the proof of Theorem \ref{thm: root n consistency for FEPLS}, we can relax the assumption that $X$ follows a Gaussian process to assuming that $X$ follows some random process with finite fourth moments, which can still guarantee the root-$n$ consistency (see Proposition 8 in \cite{cook_helland_su_2013}).

If Assumption \ref{assump: finite-rank operator} does not hold, that is, $\ran(B^*)$ is infinite-dimensional, then we cannot find a projection to a finite Hilbert space that can keep all the material part of (\ref{eq: FPELM}). Since $\HY$ being finite-dimensional necessarily implies that $B$ is finite-rank, then $\HY$ must be infinite-dimensional when Assumption \ref{assump: finite-rank operator} does not hold.  Although Assumption 1 does not hold, the conclusion of Proposition \ref{prop: FPELM and MPELM} can still hold for some $\HXmx$. We will show the existence of such a $\HXmx$ based on a \KL\ expansion such that Proposition \ref{prop: FPELM and MPELM} holds. Consider the \KL\ expansion of $\bX$: 
\begin{equation*}
    \bX(\cdot)=\suml{i=1}^\infty \psi_i b_i(\cdot),
\end{equation*}
where $\psi_i$ follows i.i.d. $N(0,\sigma^2_{\psi_i})$ and $E_{KL}=\{b_i(\cdot):i\in\Ns\}$ is an orthonormal basis of $\HX$. Then for any $\bd{b}_{m_x}$, which is a subset of $E_{KL}$, and any $\bd{c}_{m_y}$, Proposition \ref{prop: FPELM and MPELM} still holds. A proof of this result is included in Appendix A. As we increase both $m_x$ and $m_y$ with increasing $n$, then $\Bmymx$ can approximate a compact $B$, which leads to the following theorem.

\begin{theorem}
    Assume $\bd{b}_{m_x}$ is the first $m_x$ functions of $E_{KL}$, and $\bd{c}_{m_y}$ is the first $m_y$ functions of an orthonormal basis of $\HY$. Then, for a compact $B$, there exists a sequence $(m_x,m_y,n)$ such that the corresponding FEPLS estimator $\Bhatmymx^{(n)}$ is consistent for $B$, i.e.
    \begin{equation*}
        ||\Bhatmymx^{(n)}-B||_{op} \overset{p}{\to} 0,\ \text{as } n\to \infty,
    \end{equation*}
    where $m_x=m_x(n)$ and $m_y=m_y(n)$ can be viewed as functions of $n$.
    \label{thm: consistency for FEPLS under infty dim}
\end{theorem}

Theorem \ref{thm: consistency for FEPLS under infty dim} unveils the possibility of addressing infinite-dimensional $\overline{\ran}(B^*)$ by the \KL\ expansion of $\bX$. In practice, using the empirical $E_{KL}$ is not required to achieve good numerical performance, as the first step is to keep the material part, which does not require the special basis functions.


\subsection{Confidence and prediction intervals}
Now we detail how to construct pointwise confidence and prediction intervals based on the aforementioned asymptotic approximation. In this subsection, we reintroduce the intercept to account for variability in its estimation. For simplicity, we only consider functional responses in this section, though our discussion also applies for scalar responses. Let $\boldsymbol{X}_{\text{new}}$ be a new observation of $\boldsymbol{X}$. Specifically, we are interested in the confidence interval of $E(Y_{\text{new}}(t_0))$ and prediction interval for $Y_{\text{new}}(t_0)$ for $t_0\in [a,b]$. 

Let $\Bar{Y}^{(m_y)}$ denote the sample mean of $\Yvmy$. The prediction of the new outcome is $\hat{Y}_{new}(t_0)=(\Bar{Y}^{(m_y)}+\Bmmymx \Xvmx_{\text{new}})^T\boldsymbol{c}_{m_y}(t_0)$, with corresponding variance given by
\begin{equation*}
    \Var\left(\hat{Y}_{new}(t_0)\right)=\boldsymbol{c}_{m_y}(t_0)^T\Var(\Bar{Y}^{(m_y)}+\Bmmymx \Xvmx_{\text{new}})\boldsymbol{c}_{m_y}(t_0).
\end{equation*}

Next, we apply the asymptotic normality result seen in Proposition \ref{prop: asymp normality of projected MPELM} to derive an approximation of $\Var(\Bar{Y}^{(m_y)}+\Bmmymx \Xvmx_{\text{new}})$. Since $\Bar{Y}^{(m_y)}$ is independent of $\Bmmymx \Xvmx_{\text{new}}$, an asymptotic approximation is given by
\begin{equation*}
    \widehat{\Var}\left(\hat{Y}_{new}(t_0)\right)=n^{-1}\boldsymbol{c}_{m_y}(t_0)^T\left(\hat{\Sigma}_{\Yvmy} + (\Xvmx_{\text{new}})^T \hat{V}_{mpelm} \Xvmx_{\text{new}}\right)\boldsymbol{c}_{m_y}(t_0),
\end{equation*}
where $\hat{\Sigma}_{\Yvmy}$ is the sample variance of $\Yvmy$ and $\hat{V}_{mpelm}$ is the plug-in estimator of $V_{mpelm}$. The asymptotic 95\% confidence interval for $E(Y_{new}(t_0))$ is
\begin{equation*}
    \left(\hat{Y}_{new}(t_0)-z_{0.975} \widehat{\Var}\left(\hat{Y}_{new}(t_0)\right) ,\hat{Y}_{new}(t_0)+z_{0.975}\widehat{\Var}\left(\hat{Y}_{new}(t_0)\right)\right),
\end{equation*}
where $z_{0.975}$ is the 0.975 percentile of standard normal distribution.

The prediction interval for $Y_{new}(t_0)$ can be obtained by additionally included the variability of $(\evmy)^T\boldsymbol{c}_{m_y}(t_0)$. The asymptotic 95\% prediction interval can then be  
\begin{equation*}
    \begin{split}
        &\Big(\hat{Y}_{new}(t_0)-z_{0.975} \left(\widehat{\Var}\left(\hat{Y}_{new}(t_0)\right)+\boldsymbol{c}_{m_y}(t_0)^T\hat{\Sigma}_{\evmy}\boldsymbol{c}_{m_y}(t_0)\right),\\
        &\qquad\hat{Y}_{new}(t_0)+z_{0.975} \left(\widehat{\Var}\left(\hat{Y}_{new}(t_0)\right)+\boldsymbol{c}_{m_y}(t_0)^T\hat{\Sigma}_{\evmy}\boldsymbol{c}_{m_y}(t_0)\right)\Big).
    \end{split}
\end{equation*}

\section{Generalized functional envelope linear model} 
    \label{sec: GFPELM} 
    The above discussion can be generalized to a more flexible model, the generalized functional linear model (GFLM), which is applicable to other non-normal distributions for the response. 
    

    \subsection{Generalized functional linear model}
    We start with the generalized functional linear model (GFLM), which was first introduced by \cite{müller_stadtmüller_2005}, as a natural extension of the classical generalized linear model (GLM). The standard GLM utilizes a link function to connect the linear predictor and the mean of response, which allows for a wider range of response distributions. The concept of GLMs can be extended to functional data by considering a linear predictor of functions. Let $f$ denote the probability density function or probability mass function of $Y$. The GFLM can be formulated as:
    \begin{equation*}
    \begin{split}
        &f(y;\theta)=\exp((y\theta-b(\theta))/a(\phi)+c(y,\phi)),\\
        &\mu= b'(\theta),\qquad g(\mu)=\alpha+BX,\\
    \end{split}
    \label{eq: GFLM}
    \end{equation*}
    where $X\in\Hc_{\bX}$ is a Gaussian random function, $B$: $\Hc_X\to\Rs$ is a bounded linear operator, and $Y\in\Rs$ follows a distribution within the exponential family. Similar to the setting in standard GLMs (\cite{mccullagh_nelder_1989}), the link function $g$ is assumed to be monotonic and differentiable.
    

    Next, the generalized functional (predictor) envelope linear model (GFELM) with a predictor envelope $\Ec$ can be formulated as  
    \begin{equation}
        \begin{split}
            f(y;\theta)&=\exp((y\theta-b(\theta))/a(\phi)+c(y,\phi)),\qquad \mu= b'(\theta),\\
        g(\mu)&=\alpha+ \eta^* P_{\Ec}^* \bX,\qquad \SigmaX=P_{\Ec}\SigmaX P^*_{\Ec}+Q_{\Ec}\SigmaX Q^*_{\Ec}.
        \label{eq: GFELM}
        \end{split}
    \end{equation}
    For (\ref{eq: GFELM}), the regression coefficient operator $B=\eta^* P_{\Ec}^*$ is a quantity of particular interest. Let $E_x$ be a set of the eigen functions of $\SigmaX$. With $E_x$,
   (\ref{eq: GFELM}) can be rewritten as
    \begin{equation*}
        \begin{split}
            f(y;\theta)&=\exp((y\theta-b(\theta))/a(\phi)+c(y,\phi)),\qquad \mu= b'(\theta),\\
        g(\mu)&=\alpha+\suml{j\in J}\beta_jx_j,\qquad \SigmaX=\suml{j\in J}\sigma^2_{xj}e_{xj}\otimes e_{xj} + \suml{j\notin J}\sigma^2_{xj}e_{xj}\otimes e_{xj},
        \label{eq: GFELM under Ex}
        \end{split}
    \end{equation*}
    where $\bX=\suml{j\in\Ns}x_j e_{xj}$ and $\{\sigma_{xj}:j\in\Ns\}$ is the spectrum of $\SigmaX$. Now that we have formulated the modeling framework, we can discuss estimation and asymptotic properties.


    \subsection{Estimation}
    
    
    For a GFELM, since $B\bX\in\Rs$, $B$ is always finite-rank, i.e., Assumption \ref{assump: finite-rank operator} always holds. Following the idea of FPELM, the finite-dimensional projection with all material part is possible and the following proposition holds.

    \newcommand{\Bmx}{B_{m_x}}
    
    \begin{proposition}
        Suppose that $\ran(B^*)\subseteq\HXmx$. Then, $(\Xmx, Y)$ still follows a GFELM:
        \begin{equation}
            \begin{split}
             &f(y;\theta)=\exp((y\theta-b(\theta))/a(\phi)+c(y,\phi)),\\
             &\mu= b'(\theta),\qquad g(\mu)=\alpha+\Bmx\Xmx,\\
             &\Sigma_{\Xmx}=\suml{j\in J}\sigma_{xj}^2 P_{\Xmx}e_{xj}\otimes (P_{\Xmx}e_{xj}) + \suml{j\notin J}\sigma_{xj}^2P_{\Xmx}e_{xj}\otimes (P_{\Xmx}e_{xj}),
             \label{eq: projected GFELM}
            \end{split}
        \end{equation}
    where $\Bmx:\HXmx\to\Rs $ satisfies that for any $f\in\HXmx$, $\Bmx f=Bf$. With $\bd{b}_{m_x}$, (\ref{eq: projected GFELM}) can be written as a GMELM:
    \begin{equation}
            \begin{split}
             f(y;\theta)&=\exp((y\theta-b(\theta))/a(\phi)+c(y,\phi)),\qquad \mu= b'(\theta),\\
             g(\mu)&=\alpha+\Tilde{B}_{m_x}\Xvmx,\qquad \Sigma_{\Xvmx}=\suml{j\in\Ns}\sigma_{xj}^2 \evxj \evxj^T.
             \label{eq: projected GMELM}
            \end{split}
    \end{equation}
    where $\Tilde{e}_{xj}^{(m_x)}\in \Rs^{m_x}$ and $\Xvmx\in \Rs^{m_x}$ are the coordinates of $P_{\HXmx}e_{xj}$ and $\Xmx$ with respect to $\bd{b}_{m_x}$.
    \label{prop: GFELM and GMELM}
    \end{proposition}
    
    The generalized functional envelope-based PLS (GFEPLS) estimator can also be constructed based on the functional predictor envelope in GFLM. As before, we implement a two-step estimation procedure where the first step involves the coordinate estimation of $\Xvmx$ by OLS on suitable orthonormal basis functions $\bd{b}_x$. The discussion for the first step in Section \ref{subsec: estimation in FPELM} can be applied here. In the second step, we fit a GMELM on ($\Xvmx,Y$), with the method discussed in \ref{subsec: GMELM}. The prediction can be obtained by the GMELM of ($\Xvmx,Y$).

    \subsection{Theoretical properties of GFEPLS estimators}
    Similarly, we use $\Bhatmmx$ and $\Bhatmx$ to denote the GEPLS estimator and the corresponding GFEPLS estimator in the previous subsection. For simplicity, we assume that the distribution of $Y$ belongs to a natural exponential family. In this subsection we will focus on root-$n$-consistency of $\Bhatmx$.

    \begin{proposition}
        Suppose Assumption \ref{assump: finite-rank operator} holds and $\ran(B^*)\subseteq\HXmx$. Then
        \begin{equation*}
            \sqrt{n}(\Bhatmmx-\Bmmx)\overset{d}{\to} N(0, V_{gmelm}).
        \end{equation*}
        \label{prop: asymp normality of projected GMELM}
    \end{proposition}

    An expression for $V_{gmelm}$ can be found in the appendix. The connection of $\Bhatmmx$ and $\Bhatmx$ with Proposition \ref{prop: asymp normality of projected GMELM} implies the root-$n$-consistency of $\Bhatmx$.
    
    \begin{theorem}
    Suppose Assumption \ref{assump: finite-rank operator} holds and $\ran(B^*)\subseteq\HXmx$. The GFEPLS estimator, $\Bhatmx$, is root-$n$-consistent in the sense that:
    \begin{equation*}
        nE||\Bhatmx-B||^2_{op}<\infty, \text{as } n\to\infty,
    \end{equation*}
    and $\hat{\alpha}$ is also root-$n$-consistent.
    \label{thm: root n consistency for GFEPLS}
    \end{theorem}

    \section{Simulation study}
    In this section, we utilize simulations to illustrate the potential gains in prediction accuracy of the FEPLS and GFEPLS estimators compared with existing state-of-the-art methods. We consider scenarios with either 1) a function response, or 2) a categorical response. 
    
    \subsection{Functional Response}
    
    The first scenario considers a functional response $Y$, which aims to examine the numerical performance of our proposed FEPLS estimator when $\ran(B)\nsubseteq \HYmy$ and $\ran(B^*)\nsubseteq \HXmx$. The functional Hilbert spaces $\HX$ and $\HY$ are constructed to be finite-dimensional, which are spanned by a finite-dimensional subset of Fourier basis functions. To avoid degeneration to a MPELM, we choose the orthonormal basis functions formed by  spline functions, which is denoted by $\bd{b}_{m_x}$. Except for constants, any linear combinations of finite number of Fourier basis functions cannot be represented as a finite-dimensional vector in the spline function space, which ensures that the this setting does not degenerate into a finite-dimensional case. Specifically, we consider the following set of Fourier basis functions:
    \begin{equation*}
        E_{f,x} = \{e_{x1},...,e_{x,13}\}=\{1,\sqrt{2}\sin(2\pi t),\sqrt{2}\cos(2\pi t),\cdots,\sqrt{2}\cos(12\pi t)\}.
    \end{equation*}

    Let $\Hc_{X_i}=\spn(E_{f,x})$ and $\HY=\spn(E_{f,x})$, i.e. $e_{yj}=e_{xj}$ for $j=1,2,...,13$. As for $\bd{b}_{m_x}$ and $\bd{c}_{m_y}$, we utilize an orthogonalized set of natural spline functions. The predictor was generated as $X(t)=\suml{j=1}^{13}x_{j}e_{xj}(t)$, where $x_j \in \Rs^3$ follows an independent $N(0, \Sigma_{xj})$ distribution for $j = 1, 2, \ldots, 13$. For $\{\Sigma_{x1},\cdots,\Sigma_{x,13}\}$, 
    the first five are 
    \[
        \text{diag}(8,8,8),\ \text{diag}(4,2,2),\ \text{diag}(1.6,1,0.5),\ \text{diag}(7.3,7.3,3),\ \text{diag}(5.5,5.5,5.5),
    \]
    where $\text{diag}(a_1, \ldots, a_l)$ denotes a diagonal matrix with diagonal elements $a_1, \ldots, a_l$, and the rest are diagonal matrices with elements independently drawn from a uniform distribution on the interval (0.2, 0.3).
    
    

    The model expression in Proposition \ref{prop: FPELM basis expansion} was used to generate the response $Y$. The random error function $\epsilon$ was generated as $\epsilon=\suml{i=1}^{13}0.2(\Tilde{\epsilon})_ie_{yi}$, where $e_{yi}=e_{xi}$, and $\Tilde{\epsilon}_i$ follows i.i.d. $N(0,1)$ for $i=1,...,13$. The linear regression operator is $B=\suml{j=1}^{13}\suml{i=1}^{13} b_{ij}e_{y i}\otimes e_{xj}$, where $\{b_{ij}:i,j=1,...,13\}$ are all zero except 
        \[
	\begin{split}
	    &b_{22}=(-1.2,0,0)^T,\ b_{23}=(0,-0.04,0)^T,\ b_{13}=(0,-0.05,0)^T,\\
            &b_{33}=(0,0,0.03)^T,\ b_{43}=(2.4,0,-0.01)^T.
	\end{split}
	\]
    The response function was generated by $Y=\suml{i=1}^{13}\suml{j=1}^{13} b_{ij}^Tx_je_{yi}+0.2\suml{i=1}^{13}\nu_{i}e_{yi}$.  The observed data for the functional predictors and functional response were obtained by setting 16 equally spaced points over the interval $[0,1]$ as the observation points. For each predictor, the same orthonormal basis functions were obtained by orthogonalizing natural spline functions with knots $\{0,1/4,\cdots,1\}$. For the response, the orthonormal basis functions were obtained by natural spline functions with knots $\{0,1/5,\cdots,1\}$.

    We compared FEPLS to three methods, which include the full function-on-function regression model (FFFR) (\cite{Su2022}), PCR, and PLS. Similar to FEPLS, PCR and PLS can be directly performed on $(\Xvmx,\Yvmy)$. The performance of the four methods is measured by the mean square prediction error (MSPE) at observation points on a test set with a sample size of 10000. Training sample sizes ranged from 50 to 800, and for each sample size 100 replications were conducted. The simulation results are shown in Table \ref{table: sim func}, where one can see that the proposed FEPLS method outperforms the other three methods in all sample sizes, though the most significant gains are achieved when the sample size is smallest. This is the situation when dimension reduction is most impactful, and it appears that the FEPLS method is more efficient in terms of dimension reduction than PCR and PLS.

    \begin{table}[!ht]
        \centering
        \caption{\label{table: sim func} Comparisons of mean squared prediction errors.}
        \begin{tabular}{|c|c c c c c|}
	    \hline
	    \multicolumn{1}{|c|}{n} &50 &100 &200 &400&800 \\ \hline
	        FEPLS & \textbf{10.52} & \textbf{9.39} & \textbf{9.13} & \textbf{9.05} & \textbf{9.00} \\ 
                FFFR & 12.56 & 10.23 & 9.47 & 9.17 & 9.03 \\ 
                PCR & 11.47 & 9.99 & 9.39 & 9.13 & 9.01 \\ 
            PLS & 11.29 & 9.92 & 9.36 & 9.12 & 9.00 \\ 
	\hline
	\end{tabular}
    \end{table}

    \subsection{A categorical response}
    The second scenario examines a situation with a categorical response to evaluate the performance of the proposed GFEPLS estimator. For illustration, we consider $Y$ to be binary and $(Y,\bX)$ follows a functional logistic regression model given by:
    \begin{equation*}
            f(y;p)=p^y(1-p)^{1-y},\qquad log \left(\dfrac{p}{1-p} \right)=\alpha+\suml{j\in J}b_jx_j.
    \end{equation*}

    Let $\Hc_{X}=\spn(E_{f,x})$ and let the covariance operator of the predictor be $\SigmaX=\suml{j=1}^{13} \sigma^2_{x,j}e_{xj}\otimes e_{xj}$, where 
    for $\{\sigma_{x,1},\cdots,\sigma_{x,13}\}$, 
    the first five are 8, 5, 0.9, 7.3, and 2,
    and the rest are independently drawn from a uniform distribution on the interval (0.1, 0.2).


    The predictor function is generated from $X(t)= \suml{j=1}^{13}x_{j}e_{xj}(t)$, where $\{x_j\}_{j=1}^{13}$ are independently drawn from Gaussian distributions with variances $\sigma_{xj}^2$. The regression coefficient operator is $B = \suml{j=1}^{13} b_{j} 1 \otimes e_{xj}$, where ${b_{j} : j = 1, \ldots, 13}$ are all zero except $b_{2} = -0.8$ and $b_{4} = 4.4$. The linear predictor is $\mu = \log\left(\frac{p}{1-p}\right) = \suml{j=1}^{13} b_{j} x_j$. Next, the response $Y$ is generated as a Bernoulli random variable with success probability $\exp(\mu)/(1+\exp(\mu))$. Under this setting, the predictor envelope is given by $\text{span}\left(\sqrt{2} \sin{(2\pi t)}, \sqrt{2} \sin{(4\pi t)}\right)$. As before the observation points consist of 15 equally spaced points over the interval $[0, 1]$, and the basis functions $\bd{b}_{m_x}$ were obtained by orthogonalizing natural spline functions with knots $\{0,1/4,\cdots,1\}$.

    We compared GFEPLS to three other methods: GLM, elastic-net regularized generalized linear models (GLMNET), and generalized partial least squares (GPLS) (\cite{ding_gentleman_2005}). Similar to GFEPLS, these three estimators can be performed directly on $(\Xvmx, Y)$. The GLMNET method was implemented with the one-standard-error rule for selecting the tuning parameter of the penalty. The dimension for GFEPLS was selected using BIC, while the dimension for GPLS was selected via 5-fold cross-validation. The performance of the four methods is measured by the mean misclassification rate over a test set with a sample size of 10000. Training sample sizes ranges from 40 to 640, and for each sample size, 100 replications were conducted. The simulation results are presented in Table \ref{table: sim cat}. The results largely mirror those from the setting with a functional outcome, as the proposed envelope-based PLS procedure achieves the lowest misclassification rates, with the largest differences appearing for the smallest sample sizes. As the sample size increases to $n=640$, the benefits of dimension reduction are negligible and all methods perform similarly well. However, dimension reduction is hugely beneficial in the small sample size setting, and the proposed procedure is the most efficient among the dimension reduction approaches. 
    \begin{table}[!ht]
        \centering
        \caption{\label{table: sim cat} Comparisons of mean misclassification rates.}
        \begin{tabular}{|c|c c c c c|}
	    \hline
	    \multicolumn{1}{|c|}{n} &40 &80 &160 &320 &640  \\ \hline
	       GFEPLS & \textbf{0.167} & \textbf{0.143} & \textbf{0.130} & \textbf{0.124} & 0.121 \\ 
          GLM & 0.193 & 0.147 & \textbf{0.130} & \textbf{0.124} & \textbf{0.120} \\ 
          GLMNET & 0.196 & 0.157 & 0.138 & 0.129 & 0.125 \\ 
          GPLS & 0.184 & 0.151 & 0.131 & \textbf{0.124} & 0.121 \\
	\hline
	\end{tabular}
    \end{table}

    \section{Data Analysis}
    In this section we study our approach against competing methods in two real-world scenarios, again examining both a functional response and a categorical response. 
    
    \subsection{Functional Response}

\begin{figure}[t]
    \centering
    \includegraphics[width=0.75\linewidth]{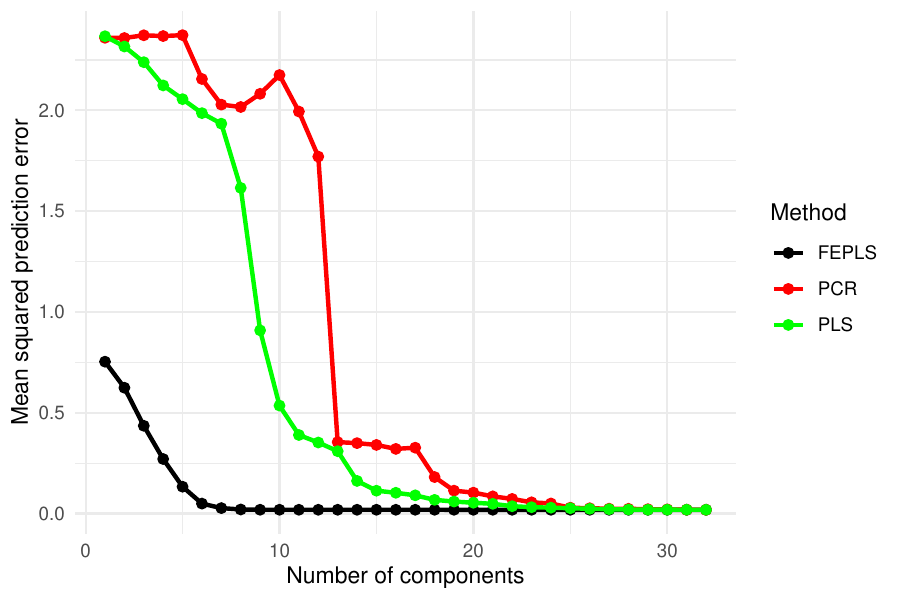} 
    \caption{Mean squared prediction error for the Hawaii ocean data across all approaches considered.}
    \label{fig: DA_functional_all}
    \end{figure}

    The Hawaii ocean data is available in the R package "FRegSigCom" and comprises five variables: salinity, potential density, temperature, oxygen, and chloropigment. These variables are measured at 2-meter intervals between 0 and 200 meters below the sea surface and can be interpreted as functions of depth. The dataset includes 116 observations, each with measurements taken at 101 evenly spaced points over the interval [0, 200]. We consider salinity as the functional response, and the remaining four variables as predictors. The dataset was also analyzed in \cite{beyaztas2021_FoFPLS} and \cite{luo_qi_2019} using function-on-function regression. Similarly to the simulation study, we apply FEPLS, PLS, and PCR to this dataset and computed the mean prediction error using five-fold cross-validation with 100 replications for each $u_x$. The results are presented in Figure \ref{fig: DA_functional_all}. The minimum mean prediction error across all methods is 0.183. However, FEPLS achieves a mean prediction error of 0.185 with only 9 components, whereas PCR and PLS require at least 31 components to reach a comparable error of 0.187. The fact that we are able to achieve such a low prediction error with only 9 components suggests that the functional predictor envelope likely has a dimensionality much smaller than 32. Additionally, while all of the approaches are able to achieve a low error rate with enough components, it is crucially important that our proposed approach is able to do so with such a significantly reduced number of components. If the sample size were smaller, the ability to retain such strong prediction with such few components would become important and would likely lead to reduced prediction error.

    \subsection{Categorical response}
    We now explore a data set on gasoline measurements (\cite{kalivas_1997}), which contains measurements of near-infrared reflectance (NIR) spectra for the octane numbers from 60 gasoline samples. The data set is available in the \emph{refund} package (\cite{goldsmithetal}). The predictors are the log(1/reflectance) at 401 wavelength ranging from 900 nm to 1700 nm in increments of 2-nm, and the response is whether the octane number is greater than 88. This dichotomization is used as gasoline grades are typically placed into one of three categories (see \url{https://www.energy.gov/}): Regular (octane number 87), Midgrade (Octane number 89-90), and Premium (Octane number 91-94). 
    
\begin{figure}[t]
    \centering
    \includegraphics[width=0.75\linewidth]{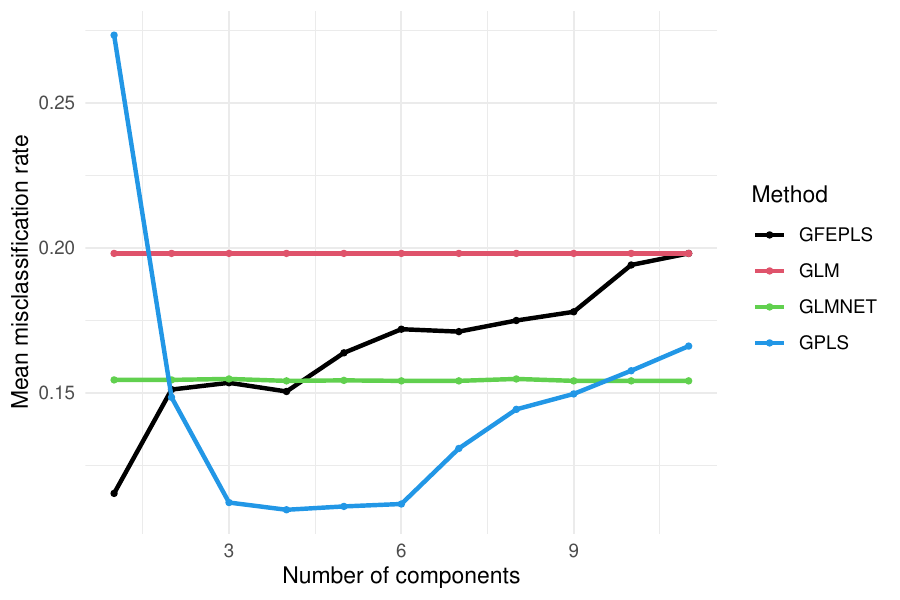} 
    \caption{Misclassification rates for the gasoline data across all estimators}
    \label{fig: DA_binary_all}
    \end{figure}

    We compared the misclassification rate of GFEPLS with the aforementioned GPLS and GLMNET. BIC was used to select the dimension of the predictor envelope, and 5-fold cross-validation was employed to select the dimension for FGPLS. The misclassification rate was estimated using 5-fold cross-validation with 100 random splits. The results are presented in Figure \ref{fig: DA_binary_all}. The misclassification rate is relatively high for both GLM and GLMNET with rates of 19.8\% and 15.4\%, respectively. The minimal misclassification rate for GFEPLS is 11.5\% and occurs at $u_x=1$, while the minimum error rate for GPLS is 11.0\% and occurs at $u_x=4$. These both significantly outperform GLM and GLMNET, with more than 25\% reduction in misclassification rates highlighting the benefits of dimension reduction in this setting. Notably again the proposed GFEPLS estimator is able to achieve sound prediction performance with a very small number of components compared to the number utilized by GPLS. 
    
    
    \begin{table}[!ht]
        \centering
        \caption{\label{table: da cat}  Minimal misclassification rates for each estimator in the gasoline data }
        \begin{tabular}{|c| c c c c|}
	    \hline
	      & GFEPLS & GLM & GLMNET & GPLS  \\
	    \hline
	    MMR  & 0.115  & 0.198 &  0.154 & 0.110 \\ 
        \hline
	\end{tabular}
     \end{table}

    \section{Discussion}

    In this paper, we first extend envelope-based PLS from the multivariate setting to the functional setting. Under a sparse structure, root-$n$ consistency of the proposed estimator is shown, paving the way to construct pointwise asymptotic confidence intervals and prediction intervals. Moreover, we develop functional envelope-based PLS in the generalized functional linear regression setting and show its root-$n$ consistency, which can be used for classification problems. To our knowledge, we are the first to show consistency in such a setting, as existing works such as \cite{delaigle_hall_2012} and \cite{preda2007pls} focus strictly on the functional linear model.

    One possible future research direction is that, following the same spirit, one can extend other envelope-based PLS methods in the multivariate setting to functional regression. For example, methods like partial partial least squares \cite{park_su_chung_2022} and sparse partial least squares \cite{zhu_2020_envelopebased} could be adapted and studied for consistency in this context. It is noteworthy that the adaptation of sparse partial least squares can also address potential high-dimensional issues brought by the first step and be applied to high-dimensional settings. Additionally, our framework opens the door to investigating functional PLS under a Bayesian perspective. Since envelope-based PLS has already been formulated within the Bayesian paradigm (see \cite{Chakraborty02072024}), these ideas can be naturally extended to the functional domain. 
    


\newcommand{\CA}{\bd{C}_{\bd{A}}}
\newcommand{\CAo}{\bd{C}_{\bd{A}_1}}

\begin{appendix}

\section{Proof}
\subsection{Proof of Proposition \ref{prop: properties of FPE}}
\begin{proof}
    For simplicity, let $\Ec$ be an abbreviation of $\penvlp$ and $\Uc=\overline{\ran}(B^*)$.
    \begin{longlist}
        \item Note that $\Ec$ is still an invariant subspace of $\SigmaX$. We only need to show $\Ec^{\perp}$ is also an invariant subspace. Let $f\in\Ec^{\perp}$. Since $f^\perp\in\Ec$, then there is $\Sc\in\text{Lat}_{\Uc}(\SigmaX)$ s.t. $f^\perp\in\Sc$. Since $\Ec\subseteq\Sc$, then $\Sc^\perp\subseteq\Ec^{\perp}$. Since $\Sc$ is a reducing subspace, then $\Sc^\perp$ is an invariant subspace. Therefore, $\SigmaX f\in\Sc^\perp$ and $\SigmaX f\in\Ec^\perp$, i.e. $\Ec^{\perp}$ is also an invariant subspace.
        
        \item Since $\Ec$ is a still reducing subspace of $\SigmaX$, then by the spectral decomposition of the compact adjoint operator or by \KL\ expansion, $P_{\Ec}\bX$ is independent of $Q_{\Ec}\bX$.
        \item Since $\Uc\subset\Ec$, then $B=BP_\Ec$. Thus, $BP_\Ec\bX$ is independent of $Q_\Ec\bX$. Note that $\epsilon$ is also independent of $Q_{\Ec}\bX$, then $Y= BP_\Ec\bX+\epsilon$ is independent of $Q_{\Ec}\bX$. 
         \item Since $I=P_\Ec+Q_{\Ec}$, then we only need to show that $Q_{\Ec}\SigmaX P^*_{\Ec}=E(P_\Ec X\otimes Q_{\Ec}X)=0$, i.e., for any $f\in\HX$ and $g\in Q_{\Ec}\HX$, $E(\inprod[\HX]{f}{(P_\Ec X\otimes Q_{\Ec}X)g})=0$.
        Using $(P_\Ec X\otimes Q_{\Ec}X)g= P_\Ec X\inprod[\HX]{Q_{\Ec}X}{g}$, we have $E(\inprod[\HX]{f}{(P_\Ec X\otimes Q_{\Ec}X)g})=E(\inprod[\HX]{f}{P_\Ec X}\inprod[\HX]{Q_{\Ec}X}{g})$=0.
        \item By the fact that $\Ec$ is still reducing subspace of $\SigmaX$, there exists $\{e_{xj}:j\in J\}$ is an orthonormal basis of $\Ec$ and $e_{xj}$ is the eigenfunction of $\SigmaX$ for $j\in J$. Next, we expand $\{e_{xj}:j\in J\}$ to an orthonormal basis $E_x=\{e_{xj}:j\in \Ns\}$ of $\HX$.
        
    \end{longlist}
\end{proof}

\subsection{Result \ref{result: infinite vector} and its proof}
We assume $\HY=\Rs$. The following result connects $\bX$ and $\SigmaX$ with an infinite-dimensional random vector and an infinite-dimensional diagonal matrix, respectively.

\begin{result}
    With $E_x$, we have:
    \begin{longlist}
        \item $\Sigma_X$ can be represented by an infinite-dimensional diagonal matrix diag$(\sigma^2_{x1},....,\sigma^2_{xj},...)$, where
        $\sigma_{xj}^2\in\Rs$ ($j\in\Ns$).
        \item $X$ can be represented by an infinite-dimensional random vector $\{x_j:x_j\in\Rs\}_{j=1}^\infty$, where $\{x_{j}:j\in\Ns\}$ follows independent Gaussian distribution with mean 0 and variance $\sigma^2_{xj}$.
        \item $B$ can be represent by an infinite-dimensional vector $\{\beta_j:\beta_j\in\Rs\}_{j=1}^\infty$.
        \item Here, for $j\in\Ns$, $\sigma^2_{xj}$ are the eigenvalues of $\SigmaX$ corresponding to $e_{xj}$, $x_j=\inprod[\Hc_{\bX}]{X}{e_{xj}}$, and $\beta_j=Be_{xj}$.
        \item Assume $E_x$ also satisfies Proposition \ref*{prop: properties of FPE}(\ref{prop item: basis}). Then $J=\{j:\beta_j\neq0\}$ and $\penvlp$ have the same cardinality.
    \end{longlist}
    \label{result: infinite vector}
\end{result}
\begin{proof}
    \begin{longlist}
        \item  Let $\SigmaX'=\suml{j\in\Ns} \sigma_{xj}^2 e_{xj}\otimes e_{xj}$. For any $f,g\in\HX$, let $f=\suml{j\in\Ns}\Tilde{f}_je_{xj}$ and $g=\suml{j\in\Ns}\Tilde{g}_je_{xj}$, where $\Tilde{f}_j=\inprod[\HX]{e_{xj}}{f}$ and $\Tilde{g}_j=\inprod[\HX]{e_{xj}}{g}$, then (\textbf{check})
        \begin{equation*}
            \begin{split}
                \inprod[\HX]{f}{(\SigmaX-\SigmaX')g} &= \inprod[\HX]{\suml{j\in\Ns}\Tilde{f}_je_{xj}}{(\SigmaX-\SigmaX')(\suml{j\in\Ns}\Tilde{g}_je_{xj})} \\
                &=0.
            \end{split}
        \end{equation*}
        It follows $\SigmaX=\SigmaX'$.
        
        \item Since $E_x$ is an orthonormal basis of $\HX$, then there exists $\{x_j:x_j\in\Rs\}_{j=1}^\infty$ such that $\bX=\suml{j\in\Rs}x_je_{xj}$. And we also have $B\bX=\suml{j\in\Ns}\beta_j x_j$
    
        Since $\bX$ follows Gaussian process with mean 0, then $x_j= \inprod[\HX]{e_{xj}}{\bX}$ follows Gaussian distribution. To show the independence, we only need to show $E x_jx_k=0$.
        \begin{equation*}
            \begin{split}
                E x_jx_k&= E \inprod[\HX]{e_{xj}}{\bX} \inprod[\HX]{e_{xk}}{\bX}\\
                &= E(\inprod[\HX]{e_{xj}}{\SigmaX e_{xk}})\\
                &=0
            \end{split}
        \end{equation*}
        
        \item By the Riesz representation theorem, there is an unique $f_B\in\HX$ such that $B\bX=\inprod[\HX]{f_{B}}{\bX}$. Under $E_x$, there exists an unique infinite-dimensional vector $\{\beta_j:\beta_j\in\Rs\}_{j=1}^\infty$, such that $f_B=\suml{i\in\Ns}\beta_{j}e_{xj}$.
        \item By the above proof, this holds.
        \item  Note that $\penvlp=\overline{\spn}(e_{xj}:j\in J)$. Therefore, $J$ and $\penvlp$ have the same cardinality.
    \end{longlist}

\end{proof}

\subsection{Proof of Proposition \ref{prop: FPELM basis expansion}}
    \begin{proof}
             For any finite-rank $B\in\Bc(\HX,\HY)$, it can be decomposed as $\suml{i\in\Ns}\suml{j\in\Ns}\beta_{ij} e_{yi}\otimes e_{xj}$.
            
            For any $i\in\Ns$, $l:\HX\to\Rs,\  l(f)=\inprod[\HY]{e_{yi}}{Bf}$ defines a bounded linear functional on $\Hc_{\bX}$. According to Riesz representation theorem, there exists $\suml{j\in\Ns}\beta_{ij}e_{xj}$ such that for any $f\in\Hc_{\bX}$, $l(f)=\inprod[\HY]{\suml{j\in\Ns}\beta_{ij}e_{xj}}{f}$. It implies that $Bf=\suml{i\in\Ns}\suml{j\in\Ns}\beta_{ij}e_{yi}\inprod[\Hc_{\bX}]{e_{xj}}{f}$, that is, $Bf=(\suml{i\in\Ns}\suml{j\in\Ns}\beta_{ij}e_{yi}\otimes e_{xj}) f$. Together with the spectral decomposition of $\SigmaX$, the (\ref{eq: FPELM}) can be written in the form given in Proposition \ref{prop: FPELM basis expansion}.
            

    \end{proof}

\subsection{Proof of Proposition \ref{prop: FPELM and MPELM}}

\begin{proof}
    \begin{longlist}
        \item[\textbf{To a projected FPELM:}] Since ${\ran}(B^*)\subseteq \HXmx$ and $\ran(B)\subseteq \HYmy$, then 
    \begin{equation*}
        E[Y|\bX]=B\bX=BP_{\HXmx}\bX=P_{\HYmy}P_{\HXmx}\bX= E[\Ymy|\Xmx].
    \end{equation*}
    And, $\epsilonmy = P_{\HYmy}\epsilon$ is still independent of $\Xmx$ and follows Gaussian process. Thus, $(\Xmx,\Ymy)$ still follow FLM. Let $\Tilde{e}_{xj}^{(m_x)}$ denote the coordinates of $P_{\HXmx}e_{xj}$ under $\bd{b}_{m_x}$. Next, the 
    \begin{equation*}
        \begin{split}
            \Sigma_{\Xmx}
            &=E(P_{\Xmx}X\otimes X P_{\Xmx}^*)\\
            &=P_{\Xmx}\SigmaX P_{\Xmx}^*\\
            &= \suml{j\in J} \sigma_{xj}^2 P_{\Xmx}e_{xj}\otimes e_{xj}P_{\Xmx}^* + \suml{j\notin J}\sigma_{xj}^2P_{\Xmx}e_{xj}\otimes e_{xj}P_{\Xmx}^*.\\
            &= \suml{j\in J} \sigma_{xj}^2 (\Tilde{e}_{xj}^T\bd{b}_{m_x})\otimes (\Tilde{e}_{xj}^T\bd{b}_{m_x}) + \suml{j\notin J}\sigma_{xj}^2(\Tilde{e}_{xj}^T\bd{b}_{m_x})\otimes (\Tilde{e}_{xj}^T\bd{b}_{m_x})\\
            &=\suml{i,k=1}^{m_x} (\suml{j\in\Ns}\sigma_{xj}^2 (\Tilde{e}_{xj})_i(\Tilde{e}_{xj})_k) b_{m_x,i}\otimes b_{m_x,k}
        \end{split}
    \end{equation*}
    \item[\textbf{To a MPELM:}]  Let $\Xvmx$ be the coordinates of $\Xmx$ given $\bd{b}_{m_x}$. Let $\Yvmy$ and $\evmy$ be the coordinates of $\Ymy$ and $\emy$, receptively. Then $\Xmx$, $\Ymy$ and $\evmy$ still follow Gaussian distribution and $\Xvmx\indep\evmy$. Since $\Xmx$ and $\Ymy$ follow multivariate linear model, $\Xmx$ and $\Ymy$ follow MPELM. By the above proof, we have 
    $\Tilde{\Sigma}_{\Xvmx}=\suml{j\in\Ns}\sigma_{xj}^2 \evxj \evxj^T$. 
    \end{longlist}
\end{proof}

\subsection{Proof of Proposition \ref{prop: asymp normality of projected MPELM}}
\begin{proof}
    Using Proposition \ref{prop: FPELM and MPELM},  the coordinates $(\Yvmy,\Xvmx)$ follow MPELM. To detail the expressions of the asymptotic variance of $\Bhatmmymx$, we assume that the coordinate form of (\ref{eq: projected MPELM}) is 
    \begin{equation*}
        \Yvmy = \bd{\eta}^T{\Phi}^T\Xvmx+\evmy,\qquad {\Sigma}_{\Xvmx}={\Phi}{\Delta}{\Phi}^T+{\Phi}_0{\Delta}_0{\Phi}_0^T,
    \end{equation*}
    where $\bd{\eta}$ is a vector and is different from $\eta$ in (\ref{eq: FPELM}), and $\Phi\in\Rs^{m_x\times q}$.
    
    The estimator of MPELM can be obtained by maximizing the log likelihood function, which is equivalent to (\ref{eq: obj func of MPELM}). Let $\hat{\Phi}$ denote the estimator obtained by (\ref{eq: obj func of MPELM}). Then the estimators of other parameters are
    \begin{equation*}
        \begin{split}
            &\hat{\Sigma}_{\Yvmy}=S_{\Yvmy},\\
        &\hat{\Delta}=\hat{\Phi}^TS_{\Xvmx}\hat{\Phi},\\
        &\hat{\Delta}_0=\hat{\Phi}^T_0S_{\Xvmx}\hat{\Phi}_0,\\
        &\hat{\bd{\eta}}=(\hat{\Phi}^TS_{\Xvmx}\hat{\Phi})^{-1}\hat{\Phi}^TS_{\Xvmx,\Yvmy},\\
        &\Bhatmmymx^T = \hat{\Phi}(\hat{\Phi}^T S_{\Xvmx}\hat{\Phi})^{-1}\hat{\Phi}^TS_{\Xvmx,\Yvmy}.
        \end{split}
    \end{equation*}

    Let $\dag$ denote Moore-Penrose generalized inverse.
    Using Proposition 9 from \cite{cook_helland_su_2013} $\sqrt{n}(\text{vec}(\Bhatmmymx)-\text{vec}(\Bmmymx))\overset{d.}{\to}N(0,V_{MPELM})$, where 
    \begin{equation*}
        \begin{split}
            V_{MPELM} = \Phi\Delta^{-1}\Phi^T \otimes \Sigma_{\evmy} + (\Phi_0\otimes \bd{\eta}^T) M^{\dag} (\Phi_0^T\otimes\bd{\eta}),
        \end{split}
    \end{equation*}
    and 
    \begin{equation*}
        M=\Delta_0\otimes \bd{\eta}\Sigma_{\evmy}\bd{\eta}^T + \Delta_0^{-1}\otimes\Delta + \Delta_0\otimes\Delta^{-1} - 2 I_{(m_x-q)q}.
\end{equation*}

\end{proof}

\subsection{Proof of Theorem \ref{thm: root n consistency for FEPLS}}

\begin{proof}
    By Proposition \ref{prop: FPELM and MPELM}, $\Bhatmmymx$ is the envelope-based estimator for $\Bmmymx$ in MPELM. 

     The Proposition \ref{prop: asymp normality of projected MPELM} shows the root-$n$-consistence of $\Bhatmmymx$. It implies that $nE||\Bhatmmymx-\Bmmymx||^2_F<\infty$ as $n\to\infty$, where $||\cdot||$ is Frobenius norm. 
    Proposition \ref{prop: FPELM and MPELM} builds an isomorphism between projected FPELM and MPELM. The linear operator in (\ref{eq: projected FPELM mx my}) corresponding to $\Bmmymx$ is $\Bmymx$. Since $\Bhatmymx-B=\Bhatmymx-\Bmymx P_{\HXmx}=(\Bhatmymx-\Bmymx)P_{\HXmx}$, by the definition of operator norm, $||\Bhatmymx-B||_{op}= ||\Bhatmymx-\Bmymx||_{op}$, where, for the right hand right, we consider $\Bhatmymx:\HXmx \to\HYmy$. 

    Any linear operator in $\Bc(\HXmx,\HYmy)$ can be written as a $m_y\times m_x$ matrix and the operator norm is equal to the spectral norm, i.e., $||\Bhatmymx-\Bmymx||_{op}=||\Bhatmmymx-\Bmmymx||_{sp}$, where $||\cdot||_{sp}$ denotes the spectral norm. 

    Since $||\Bhatmmymx-\Bmmymx||_{sp}\leq ||\Bhatmmymx-\Bmmymx||_{F}$, then Theorem \ref{thm: root n consistency for FEPLS} holds.

\end{proof}

\subsection{Proof of Proposition \ref{prop: FPELM basis expansion} for \KL\ expansion}
\begin{proof}
    Since $P_{\HXmx}X$ is independent of $Q_{\HXmx}X$, thus ($\Xmx,\Ymy$) still follow FPELM but with different functional predictor envelope. Therefore, Proposition \ref{prop: FPELM and MPELM} still holds, but $\Bmymx P_{\HXmx}$ is not equal to $B$.
\end{proof}

\subsection{Proof of Theorem \ref{thm: consistency for FEPLS under infty dim}}

\begin{proof}
            At the first, we show that $B$ can be approximated by $P_{\HYmy}BP_{\HXmx}$. Since $B$ is compact and $B^*$ is also compact (Theorem 4.4 from Chapter 2 of \cite{conway}), then by the Corollary 4.5 from same chapter, we can show $||P_{\HYmy}B-B||_{op}\to0$, as $m_y\to\infty$ and $||P_{\HXmx}B^*-B^*||_{op}=||BP_{\HXmx}^*-B||_{op}=||BP_{\HXmx}-B||_{op}\to0$ (the projection is self-adjoint), as $m_x\to\infty$. It follows that $||P_{\HYmy}BP_{\HXmx}-B||_{op}\to0$, as $m_y\to\infty$ and as $m_x\to\infty$ (The order of the limits does not matter). This means that $B$ can be approximated by $P_{\HYmy}BP_{\HXmx}$.

            Second, given fixed $m_x$ and, we show that $||\Bhatmymx^{(n)}-B||_{op} \overset{p}{\to} 0\ \text{as } n\to \infty$. We only need to show that 
            ($\Xvmx,\Yvmy$) also follows a FPELM. Since $(X,Y)$ follows FPELM, then
            \begin{equation*}
                \begin{split}
                    \Ymy &= P_{\HYmy}(B\bX+\epsilon)\\
                         &= P_{\HYmy}B(P_{\HXmx}+Q_{\HXmx})\bX+P_{\HYmy}\epsilon\\
                         &=P_{\HYmy}BP_{\HXmx}\bX + P_{\HYmy}BQ_{\HXmx}\bX+P_{\HYmy}\epsilon.
                \end{split}
            \end{equation*} 
            
            Note that $(Q_{\HXmx}\bX,P_{\HYmy}\epsilon)$ is independent of $P_{\HXmx}\bX$ and $P_{\HYmy}BQ_{\HXmx}\bX+P_{\HYmy}\epsilon$ still follows Gaussian process. Let $\emymx$ denote $P_{\HYmy}BQ_{\HXmx}\bX+P_{\HYmy}\epsilon$. By Theorem \ref{thm: root n consistency for FEPLS}, $\Bhatmymx^{(n)}$ is consistent with $\Bmymx$.


            Consider $m_x=m_y$ and for each fixed $m_x=m_y$, choose $n(m_x)$ such that for any $n\geq n(m_x)$,  $P(||\Bhatmymx^{(n)}-\Bmymx||_{op}>1/m_x)<1/m_x$ and $n(m_x)>n(m_x-1)$ hold. Then for $n$ between $n(i)$ to $n(i+1)$, we let $m_x=m_y$ follow:
            \[
            \begin{matrix}
                \{(m_x = m_y = 1, n = 1),&\cdots,&(m_x = m_y = 1, n = n(2)),&\cdots &(m_x = m_y = 2, n = n(2)+1),&\cdots,\\
                (m_x = m_y = 2, n = n(3)),&\cdots,&(m_x = m_y = i, n = n(i)+1),&\cdots,&(m_x = m_y = i, n = n(i+1)),&\cdots.\}
            \end{matrix}
            \]

            Finally, we finish our proof by showing that for any $\delta>0$ and for any $\tau>0$, there exists $n_1$ such that for any $n>n_1$,
            $P(||\Bhatmymx^{(n)}-B||_{op}>\delta)<\tau$.  Since
            $P(||\Bhatmymx^{(n)}-\Bmymx||_{op}>1/m_x)<1/m_x$ for any $m_x>1$, let $n_1$ satisfy that 
            $n_1>n(m'_{x})$ and 
            $m'_x>\max(1/\delta,1/\tau)$.

\end{proof}

\subsection{Proof of Proposition \ref{prop: GFELM and GMELM}}

    \begin{proof}
        \begin{longlist}
            \item[\textbf{To a projected GFELM:}] Since $\ran(B^*)\subseteq \HXmx$, then $B\bX=BP_{\HXmx}\bX={B}_{m_x}\Xmx$.  Thus, $(\Xmx,Y)$ still follows a GFELM with the same link function, the same cumulant function and the same exponential family. Let $\Tilde{e}_{xj}^{(m_x)}$ denote the coordinates of $P_{\HXmx}e_{xj}$ under $\bd{b}_{m_x}$. Referring to the proof of Proposition \ref{prop: FPELM and MPELM}, we still have
            
            \[
            \begin{split}
                \Sigma_{\Xmx}
                &= \suml{j\in J} \sigma_{xj}^2 P_{\Xmx}e_{xj}\otimes e_{xj}P_{\Xmx}^* + \suml{j\notin J}\sigma_{xj}^2P_{\Xmx}e_{xj}\otimes e_{xj}P_{\Xmx}^*\\
                &=\suml{i,k=1}^{m_x} (\suml{j\in\Ns}\sigma_{xj}^2 (\Tilde{e}_{xj})_i(\Tilde{e}_{xj})_k) b_{m_x,i}\otimes b_{m_x,k}.
            \end{split}
            \]
            \item[\textbf{To a GMELM:}]  Let $\Xvmx$ be the coordinates of $\Xmx$ given $\bd{b}_{m_x}$. 
            Since $\HXmx$ is finite-dimensional, given $\bd{b}_{m_x}$, there exists an unique matrix $\Tilde{B}_{m_x}in\Rs^{1\times m_x}$ such that ${B}_{m_x}\Xmx=\Tilde{B}_{m_x}\Xvmx$. Then, $g(\mu)=\alpha+\Tilde{B}_{m_x}\Xvmx$. It implies that $(\Xvmx,Y)$ follow the GMELM with an covariance matrix $\SigmaX=\suml{j\in\Ns}\sigma_{xj}^2 \evxj \evxj^T$.

        \end{longlist}

    \end{proof}
\subsection{Proof of Proposition \ref{prop: asymp normality of projected GMELM}}
\begin{proof}
    By Proposition \ref{prop: GFELM and GMELM}, $(Y,\Xvmx)$ follow GMELM. We assume that the coordinate form of (\ref{eq: projected GMELM}) is
    \begin{equation*}
    \begin{split}
        &log(f(Y\mid\theta))=Y\theta-b(\theta)+c(y),\qquad \theta(\mu)=(b')^{-1}(g^{-1}(\mu))\\
        &\mu=\alpha+\bd{\eta}^T{\Phi}^T\Xvmx,\qquad\Sigma_{\Xvmx}={\Phi}\Delta{\Phi}^T+{\Phi}_0\Delta_0{\Phi}_0^T,
    \end{split}
    \end{equation*}
    where $\bd{\eta}$ is a vector and is different from $\eta$ in (\ref{eq: FPELM}), and $\Phi\in\Rs^{m_x\times q}$.

As discussed in \cite{cook_2015_foundations}, the estimator of the parameters can be obtained by maximizing the log likelihood function. After simplification, this is equivalent to minimize (\ref{eq: obj func of GMELM}) and the optimization of (\ref{eq: obj func of GMELM}) can be solved by the alternating optimization described in Appendix \ref{append: alternating optimization}. Let $\hat{\alpha}$, $\hat{\Phi}$ and $\hat{\bd{\eta}}$ denote the obtained estimator by optimizing (\ref{eq: obj func of GMELM}). According to Proposition 5 from \cite{cook_2015_foundations}, $\sqrt{n}(\Bhatmmx-\Tilde{B}_{m_x})\overset{d}{\to}N(0,V_{gmelm})$, where
\begin{equation*}
    V_{gmelm}= P_{\Phi}\Tilde{V}P_{\Phi} + (\bd{\eta}^T\otimes\Phi_0)M^{-1} (\bd{\eta}\otimes\Phi_0^T),
\end{equation*}
$\Tilde{V}$ denotes the asymptotic variance of maximum likelihood estimator of GLM corresponding to (\ref{eq: projected GMELM}), and
\begin{equation*}
    M=\Delta_0\otimes \bd{\eta}\Tilde{V}\bd{\eta}^T + \Delta_0^{-1}\otimes\Delta + \Delta_0\otimes\Delta^{-1} - 2 I_{(m_x-q)q}.
\end{equation*}
\end{proof}

\subsection{Proof of Theorem \ref{thm: root n consistency for GFEPLS}}
    \begin{proof}
        Using Propositions \ref{prop: GFELM and GMELM} and \ref{prop: asymp normality of projected GMELM}, $\Bhatmmx$ is the envelope-based estimator for ${B}_{m_x}$ in GMELM and is also root-$n$-consistent ($\hat{\alpha}$ is also). It follows that $nE||\Bhatmmx-\Tilde{B}_{m_x}||^2_F<\infty$ as $n\to\infty$, where $||\cdot||$ is Frobenius norm. Since $||\Bhatmx-B||^2_{op}=||\Bhatmmx-\Tilde{B}_{m_x}||^2_{sp}$ (see the proof of Theorem \ref{thm: root n consistency for FEPLS}), then $nE||\Bhatmx-B||^2_{op}<\infty,$ as $n\to\infty$.
    \end{proof}
\section{Two-step estimation}
Here, we detail a two-step estimation procedure. Although we use OLS to estimate the coordinates for illustration, other methods, such as smoothing splines, penalized splines, and others, could also be applicable in this framework. 


\subsection{Coordinate estimation}
The first step is to obtain the coordinate of the predictor and the response if the response is also a function. For simplicity, we show only how to obtain the  OLS coordinate estimate for the predictors. Let $\bd{b}=(b_{1},,...,b_{m_x})$ be an orthonormal basis for $\HXmx$ and $\bd{t}_x=(t_1,t_2,\dots,t_k)$ be some observed points for $\bX$. We denote $(b_{j}(t_i))_{k\times m_x}\in\Rs^{k\times m_x}$ by $\bd{b}(\bd{t}_x)$ and denote $(\bX(t_1),\dots,\bX(t_k))^T\in\Rs_{k\times p}$ by $\bd{X}(\bd{t}_x)$. The coordinate $\Xvmx$ can be estimated by $(\bd{b}(\bd{t}_x)^T\bd{b}(\bd{t}_x))^{-1}\bd{b}(\bd{t}_x)^T\bd{X}(\bd{t}_x)$. Let $\bd{c}=(c_1,\dots,c_{m_y})$ be an orthonormal basis for $\HYmy$. Similarly, this can be applied to the response.

Note that the above coordinate estimation still works when $\bd{t}_x$ differs across individuals. When 
$(\bd{b}(\bd{t}_x)^T\bd{b}(\bd{t}_x))$ is not full rank, e.g., $k<m_x$, one could also use penalized regression, such as LASSO or ridge regression. 


\subsection{Envelope estimation and prediction}
The second step is to fit a MPELM or GMELM on the estimated coordinates.

\subsubsection{MPELM}
For illustration, we consider function response. Let $\Xvmx$ and $\Yvmy$ denote the estimated coordinates. Then, we can fit a MPELM for $\Xvmx$ and $\Yvmy$ to obtain th MPELM estimator $\Bhatmmymx$ for $\Bmmymx$. Then the prediction for a new predictor $\bX_{new}$ can be: first obtain the coordinate estimation of it, $\Tilde{\bX}_{new}^{(m_y)}$, with respect to the chosen basis $\bd{b}$. Then the predicted $\Yvmy$ is simply $\Bhatmmymx\Tilde{\bX}_{new}^{(m_y)}$ and the predicted $\Ymy$ is $\bd{c}^T\Bhatmmymx\Tilde{\bX}_{new}^{(m_y)}$.

\subsubsection{GMELM}
We fit a GMELM for $\Xvmx$ and $Y$ to obtain GFEPLS estimators: $\Bhatmmx$ for $\Bmmx$ and $\hat{\alpha}$ for $\alpha$. Then the prediction for a new predictor $\bX_{new}$ can be: first obtain the coordinate estimation of it, $\Tilde{\bX}_{new}^{(m_y)}$, with respect to the chosen basis $\bd{b}$. Next, we have $g(\hat{\mu})=\hat{\alpha}+\Bhatmmx\Tilde{\bX}_{new}^{(m_y)}$ and follow the standard GLM perdition.

\section{Alternating optimization for GMELM \label{append: alternating optimization}}
In here, we detail how to implement the alternating optimization mentioned in Section \ref{subsec: GMELM}. For convenience, We restate the objection function (\ref{eq: obj func of GMELM}) of envelope-based estimator here:
\begin{equation}
    \begin{split}
        L_n(\alpha,\bd{\eta},\bd{\Gamma})&=-\frac{2}{n} \suml{i=1}^n D(\alpha+\bd{\eta}^T\bd{\Gamma}^T\bd{X}_i)+\log|\bd{\Gamma}^T\bd{S_X}\bd{\Gamma}|+\log|\bd{\Gamma}^T\bd{S_X}^{-1}\bd{\Gamma}|,
    \end{split}
    \label{eq: obj func of GMELM appendix}
\end{equation}
where $D(\cdot)=\mathcal{C}(\theta(\cdot))$ and $\mathcal{C}(\nu)=Y\nu-b(\nu)$.

The alternating optimization includes two steps: First, given fixed $\bd{\Gamma}$, (\ref{eq: obj func of GMELM appendix}) degenerate to the objective function of GLM, which can be solved via the Fisher scoring method or any other methods used in GLM. Second, given $\alpha$ and $\bd{\eta}$, this involves an optimization over the Grassmann manifold. We can adapt the scheme proposed by \cite{cook_forzani_su_2016} to this situation. We iterate through these two steps alternately until convergence is achieved.

For the illustration purpose, we only show the details of the alternating optimization steps for GMELM within a logistic regression. The first part of alternating optimization is straightforward, so I only provide the details for the second part. For a logistic regression with a canonical link, (\ref{eq: obj func of GMELM appendix}) can be rewritten as:

\begin{equation*}
    \begin{split}
        L_n(\alpha,\bd{\eta},\bd{\Gamma})&=-\frac{2}{n} \suml{i=1}^n \left(\zeta_i y_i - \log(1+\exp(\zeta_i)\right)+\log|\bd{\Gamma}^T\bd{S_X}\bd{\Gamma}|+\log|\bd{\Gamma}^T\bd{S_X}^{-1}\bd{\Gamma}|,
    \end{split}
    \label{eq: obj func of LE}
\end{equation*}
where $\zeta_i=\alpha+\bd{\eta}^T\bd{\Gamma}^T\bd{X_i}$.

\subsection{\texorpdfstring{Optimization over \(\alpha\) and \(\bd{\eta}\) given \(\bd{\Gamma}\)}{Optimization over alpha and eta given Gamma}}

Given $\bd{\Gamma}$, the objective function is the same as the objective function of when we fit a logistic regression for $Y$ and $\bd{\Gamma}\bd{X}$. The commonly used optimization methods like Fisher scoring method can be applied. 

\subsection{\texorpdfstring{Optimization over \(\bd{\Gamma}\) given \(\alpha\) and \(\bd{\eta}\)}{Optimization over Gamma given alpha and eta}}

Given $\alpha$ and $\bd{\eta}$, the objective function for the optimization over $\bd{\Gamma}$ can be rewritten as:

\begin{equation*}
    \begin{split}
        F(\bd{G})&=-\frac{2}{n} \suml{i=1}^n (\hat{\mu}_i y_i - \log(1+\exp(\hat{\mu}_i))+\log|\bd{G}^T\bd{S_X}\bd{G}|+\log|\bd{G}^T\bd{S}^{-1}_{\bd{X}}\bd{G}|,
    \end{split}
\end{equation*}
where $\hat{\mu}_i=\hat{\alpha}+\hat{\bd{\eta}}^T\bd{\Gamma}^T\bd{X_i}$

Without loss of generality(\cite{cook_forzani_su_2016}), $G$ can be reparameterized as
\begin{equation*}
    \begin{split}
    \bd{G}&=\begin{pmatrix}\bd{G}_1\\\bd{G}_2\end{pmatrix}=\begin{pmatrix}\bd{I}_u\\ \bd{A}\end{pmatrix}\bd{G}_1=\bd{C_A}\bd{G}_1
    \end{split}
\end{equation*}
where $\bd{G}_1\in\Rs^{u\times u}$ is an invertible matrix, $\bd{A}=\bd{G}_2\bd{G}^{-1}_1\in\Rs^{(m_x-u)\times u}$ is an unconstrained matrix, and $\bd{C_A}=(\bd{I}_u,\bd{A}^T)^T$.

Since $\bd{G}^T\bd{G}=\bd{I}_u$, then we have
\begin{equation}
    \bd{G}_1\bd{G}_1^T=(\bd{C}_{\bd{A}}^T\bd{C_A})^{-1},
    \label{eq: reparameterization constraint}
\end{equation}

Given $\bd{A}$, there are infinite many $\bd{G}_1$ such that (\ref{eq: reparameterization constraint}) holds. Note that (\ref{eq: reparameterization constraint}) also implies that $\spn(\bd{G})$ is equivalent to $\spn{(\bd{C_A})}$, which lead to the same envelope. Without loss of generality, we can take $\bd{G}_1$ as $(\bd{C}^T_{\bd{A}}\bd{C_A})^{-\frac{1}{2}}$, which is a special case of $\bd{G}_1$.

With $\bd{G}_1=(\bd{C}^T_{\bd{A}}\bd{C_A})^{-\frac{1}{2}}$, $F(\bd{G})$ can be expressed as a function of $\bd{A}$:
\begin{equation}
    \begin{split}
        F(\bd{A})&=-\frac{2}{n} \suml{i=1}^n (\hat{\mu}_i y_i - \log(1+\exp(\hat{\mu}_i)) +\log|\bd{C}_{\bd{A}}^T\bd{S_X}\bd{C}_{\bd{A}}|+ \\ &\qquad \log|\bd{C}_{\bd{A}}^T\bd{S}^{-1}_{\bd{X}}\bd{C}_{\bd{A}}|-2\log|\bd{C}_{\bd{A}}^T\bd{C}_{\bd{A}}|,\\
    \end{split}
    \label{eq: reparameterized objective function of GMELM}
\end{equation}
where $\hat{\mu}_i=\hat{\alpha}+\hat{\bd{\eta}}^T(\bd{C}^T_{\bd{A}}\bd{C_A})^{-\frac{1}{2}}\bd{X_i}$. Note that $A$ is an unconstrained matrix and after reparameterization, (\ref{eq: obj func of GMELM appendix}) is transferred from a nonconvex optimization problem over the Grassmann manifold into a optimization problem over $\Rs^{(m_x-u)\times u}$

When $u=1$, (\ref{eq: reparameterized objective function of GMELM}) can be directly optimized over a row vector $\bd{A}$. For $1<u< p$, the optimization of (\ref{eq: reparameterized objective function of GMELM}) can be done through iterative row optimizations of $\bd{A}$. We provide the exact form of the objective function and gradient function for optimizing over the last row, and the approach for the other rows is analogous. 

\subsubsection{Row objective function}

Let $\bd{M}$ denote $\bd{S}_{\bd{X}}$ and $\bd{V}$ denote $\bd{S}_{\bd{X}}^{-1}$. Consider $\bd{M}$ divided into four blocks:
\[
\begin{pmatrix}
    M_{11}&\bd{M}_{12}\\
    \bd{M}_{12}^T&\bd{M}_{22}
\end{pmatrix},
\]
where $M_{11}\in\Rs$, $\bd{M}_{12}\in\Rs^{1\times(m_x-1)}$, and $\bd{M}_{22}\in \Rs^{(m_x-1)\times(m_x-1)} $. Similarly, the matrix $\bd{V}$ can also be divided into four blocks.

Next, the row objective function can be written as follows:

\begin{equation*}
    \begin{split}
        F(\bd{a}|\bd{A}_1)&=F_1(\bd{a}|\bd{A}_1)+ \log(1+M_{22}(\bd{a}+\bd{C}_{\bd{A}_1}^T\bd{M}_{12}/M_{22})^T\bd{W}_1^{-1}(\bd{a}+\bd{C}_{\bd{A}_1}^T\bd{M}_{12}/M_{22}))+\\&\qquad\log(1+V_{22}(\bd{a}+\bd{C}_{\bd{A}_1}^T\bd{V}_{12}/V_{22})^T\bd{W}_2^{-1}(\bd{a}+\bd{C}_{\bd{A}_1}^T\bd{V}_{12}/V_{22}))-\\&\qquad 2\log(1+\bd{a}^T(\bd{C}_{\bd{A}_1}^T\bd{C}_{\bd{A}_1})^{-1}\bd{a})\\
        &=F_1(\bd{a}|\bd{A}_1)+F_2(\bd{a}|\bd{A}_1),
    \end{split}
\end{equation*}
where $F_1(\bd{a}|\bd{A}_1)=-\frac{2}{n} \suml{i=1}^n (\hat{\mu}_i y_i - \log(1+\exp(\hat{\mu}_i))$, $\bd{C}_{\bd{A}_1}=(\bd{I}_u,\bd{A}_1^T)^T$, $\bd{A}=(\bd{A}_1^T,\bd{a})^T$, $\bd{W}_1=\bd{C}_{\bd{A}_1}^T(\bd{M}_{11}-\bd{M}_{12}\bd{M}_{12}^TM_{22}^{-1})\bd{C}_{\bd{A}_1}$, and $\bd{W}_2=\bd{C}_{\bd{A}_1}^T(\bd{V}_{11}-\bd{V}_{12}\bd{V}_{12}^TV_{22}^{-1})\bd{C}_{\bd{A}_1}$.

\subsubsection{Row gradient function}
Next, we derive the closed form of the gradient with respect to $\bd{a}$. Using (\ref{eq: reparameterization constraint}), one can have
\begin{equation*}
    \begin{split}
        \dfrac{vec(d\bd{G}_1)}{vec(d\bd{a})}&=-(\bd{G}_1\otimes\bd{I}_u+\bd{I}_u\otimes\bd{G}_1)^{-1}((\bd{I}_u+\bd{A}^T\bd{A})^{-1}\otimes(\bd{I}_u+\bd{A}^T\bd{A})^{-1}) (\bd{a}\otimes\bd{I}_u+\bd{I}_u\otimes\bd{a}).
    \end{split}
\end{equation*}

Let $\hat{\bd{\mu}}=\Hat{\alpha}\bd{1}_n+\bd{X}\CA\bd{G}_1\hat{\bd{\eta}}$, $\bd{Y}=(y_1,\cdots,y_n)^T$, $b(\bd{\mu})=(\log(1+\exp(\zeta_1)),\cdots,\log(1+\exp(\zeta_n)))$, and $b'(\bd{\mu})=(1/(1+\exp(-\zeta_1)),\cdots,1/(1+\exp(-\zeta_n)))$. Then, one can have
\begin{equation*}
    \begin{split}
        \dfrac{vec(d\hat{\bd{\mu}})}{vec(d\bd{a})}&=(\bd{\eta}^T\otimes\bd{X}\CA)\dfrac{vec(d\bd{G}_1)}{vec(d\bd{a})}+(\bd{\eta}^T\bd{G}_1^T\otimes\bd{X})\dfrac{vec(d\CA)}{vec(d\bd{a})},
    \end{split}
\end{equation*}
where 
\begin{equation*}
    \dfrac{vec(d\CA)}{vec(d\bd{a})}=\begin{pmatrix}
0 & 0 & \cdots & 0\\
\vdots& \vdots  &\ddots & \vdots  \\
0 & 0 & \cdots &0\\
1_{(pth)} & 0 &\cdots &0\\
0 & 0 & \cdots &0\\
\vdots  & \vdots  & \ddots & \vdots  \\
0 & 0 & \cdots &0\\
0 & 0 & \cdots & 1_{(u*pth)}
\end{pmatrix}.
\end{equation*}

The gradient of $F_1$ w.r.t $\bd{a}$ is 
\begin{equation*}
    \begin{split}
        \dfrac{dF_1}{d\bd{a}}&=-\dfrac{2}{n}((\dfrac{vec(d\hat{\bd{\mu}})}{vec(d\bd{a})})^T\bd{Y}-(\dfrac{vec(d\hat{\bd{\mu}})}{vec(d\bd{a})})^Tb'(\bd{\mu})).
    \end{split}
\end{equation*}

The gradient of $F_2$ w.r.t $\bd{a}$ is
\begin{equation*}
    \begin{split}
        \dfrac{dF_2}{d\bd{a}}&=\dfrac{2M_{22}\bd{W}_1^{-1}(\bd{a}+\bd{C}_{\bd{A}_1}^T\bd{M}_{12}/M_{22})}{1+M_{22}(\bd{a}+\bd{C}_{\bd{A}_1}^T\bd{M}_{12}/M_{22})^T\bd{W}_1^{-1}(\bd{a}+\bd{C}_{\bd{A}_1}^T\bd{M}_{12}/M_{22})}+\\
        &\quad\dfrac{2V_{22}\bd{W}_2^{-1}(\bd{a}+\bd{C}_{\bd{A}_1}^T\bd{V}_{12}/V_{22})}{1+V_{22}(\bd{a}+\bd{C}_{\bd{A}_1}^T\bd{V}_{12}/V_{22})^T\bd{W}_2^{-1}(\bd{a}+\bd{C}_{\bd{A}_1}^T\bd{V}_{12}/V_{22})}-\\ &\quad\dfrac{4(\bd{C}_{\bd{A}_1}^T\bd{C}_{\bd{A}_1})^{-1}\bd{a}}{1+\bd{a}^T(\bd{C}_{\bd{A}_1}^T\bd{C}_{\bd{A}_1})^{-1}\bd{a}}.
    \end{split}
\end{equation*}

With the closed form of gradient, then we can apply the existing methods like the L-BFGS method to optimized the row objective function.

\section{Additional simulations for a multivariate response}
    
    In here, we consider a 4-dimensional vector response, i.e., $\HY=\Rs^4$. Let $B=\suml{j=1}^{13}\suml{i=1}^{4}b_{ij}e_{y i}\otimes e_{xj}$, where ${b_{ij} = (\Tilde{B})_{ij} : i = 1, \ldots, 4, \text{ and } j = 1, \ldots, 13}$ are all zero except
    \begin{equation*}
        \begin{split}
	    &b_{12}=-1.2,\ b_{13}=-0.4,\\
            &b_{22}=0.4,\ b_{23}=0.6,\\
            &b_{32}=0.3,\ b_{33}=-0.4,\\
            &b_{42}=-1.2,\ b_{43}=0.3.
	\end{split}
    \end{equation*}

    The random error term is generated as $e=\suml{i=1}^{4}2\Tilde{e}_ie_{y i}$, where $e_{y 1}=(1,0,0,0)^T$, $e_{y 2}=(0,1,0,0)^T$, $e_{y 3}=(0,0,1,0)^T$, $e_{y 4}=(0,0,0,1)^T$, and $\{\Tilde{e}_i\}_{i=1}^{4}$ follows independent standard Gaussian distribution. The covariance operator is defined as $\SigmaX=\suml{j=1}^{13} \sigma^2_{x,j}e_{xj}\otimes e_{xj}$, where 
    \begin{equation*}
        \{\sigma_{x1}^2,\cdots,\sigma_{x,13}^2\}=\{8,2,0.08,12,5.5,\text{runif}(8,2,3)  \}.
    \end{equation*}
    Here, $\text{runif}(m,b,c)$ represents $m$ randomly generated numbers independently following a uniform distribution on the interval [b, c].

    The predictor function is generated as $X(t)= \suml{j=1}^{13}x_{j}e_{xj}(t)$, where $\{x_j\}_{j=1}^{13}$ are independent Gaussian random variables with variances $\sigma_{xj}^2$. Under this setting, the predictor envelope is $\text{span}(\sqrt{2}\sin(2\pi t),\sqrt{2}\cos(2\pi t))$.

    The observation points are 13 equally spaced points over the interval $[0,1]$, that is, $\{0,1/12,...,1\}$. The basis functions $\bd{b}_{m_x}$ used for the coordinate estimation of $X$ were obtained by orthogonalizing natural spline functions with equally spaced knots at $\{0,1/5,\cdots,1\}$. We compared FEPLS with ordinary least squares (OLS), principal component regression (PCR), and partial least squares (PLS). Similar to FEPLS, OLS, PCR and PLS can be performed directly on ($\Xvmx$,Y). The dimension for FEPLS was selected using BIC, while the dimensions for PCR and PLS were chosen via 5-fold cross-validation. The performance of the four methods was assessed based on the mean squared prediction error (MSPE) over a test set with a sample size of 1000. Sample sizes range from 25 to 400, and for each sample size, 100 replications were conducted. The simulation results are presented in Table \ref{table: sim scalar}.

    \begin{table}[!ht]
        \centering
        \caption{\label{table: sim scalar} MSPE (A vector response)}
        \begin{tabular}{|c|c c c c c|}
	    \hline
	    \multicolumn{1}{|c|}{n} &25 &50 &100 &200 &400 \\ \hline
	        FEPLS & 3.25 & 2.76 & 2.66 & 2.61 & 2.59 \\
	        OLS  & 4.10 & 3.12 & 2.83 & 2.68 & 2.62 \\
	        PCR  & 3.68 & 3.00 & 2.78 & 2.67 & 2.62 \\
	        PLS  & 3.63 & 2.98 & 2.77 & 2.66 & 2.62 \\
	\hline
	\end{tabular}
    \end{table}


     Table \ref{table: sim scalar} shows that when sample sizes are small, FEPLS significantly outperforms all other methods. For instance, when sample size $n=25$, it reduces MSPE of sFOLS by 21\%, MSPE of PCR by 12\%, and MSPE of PLS by 11\%. As the sample size increases, these differences diminish, as all estimators converge consistently to the projected FLM. The poor performance of OLS may be attributed to the large condition number of $\Tilde{\Sigma}_{\Xvmx}$. Unlike FEPLS, neither PCR nor PLS successfully identified the relevant material part, resulting lower prediction accuracy.

\end{appendix}
\bibliographystyle{imsart-number} 
\bibliography{my}       


\end{document}